\documentclass[aps,prb,twocolumn,superscriptaddress]{revtex4-1}
\usepackage{graphicx}
\usepackage{MnSymbol}
\usepackage{todonotes}
\begin{document}

%Title of paper
\title{Magnetically polarized Ir dopant atoms in superconducting Ba(Fe$_{1-x}$Ir$_x$)$_2$As$_2$}

\author{M. P. M. Dean}
\email{mdean@bnl.gov}
\affiliation{Department of Condensed Matter Physics and Materials Science,Brookhaven National Laboratory, Upton, New York 11973, USA}

\author{M. G. Kim}
\email{mgkim@iastate.edu}
\author{A. Kreyssig}
\affiliation{Ames Laboratory, US Department of Energy and Department of Physics and Astronomy, Iowa State University, Ames, Iowa 50011, USA}

\author{J. W. Kim}
\affiliation{Advanced Photon Source, Argonne National Laboratory, Argonne, Illinois 60439, USA}

\author{X. Liu}
\affiliation{Department of Condensed Matter Physics and Materials Science,Brookhaven National Laboratory, Upton, New York 11973, USA}

\author{P. J Ryan}
\affiliation{Advanced Photon Source, Argonne National Laboratory, Argonne, Illinois 60439, USA}

\author{A. Thaler}
\author{S. L. Bud'ko}
\author{W. Strassheim}
\author{P. C. Canfield}
\affiliation{Ames Laboratory, US Department of Energy and Department of Physics and Astronomy, Iowa State University, Ames, Iowa 50011, USA}

\author{J. P. Hill}
\affiliation{Department of Condensed Matter Physics and Materials Science,Brookhaven National Laboratory, Upton, New York 11973, USA}
\author{A. I. Goldman}
\affiliation{Ames Laboratory, US Department of Energy and Department of Physics and Astronomy, Iowa State University, Ames, Iowa 50011, USA}

% user macros
\newcommand{\sample}{ Ba(Fe$_{0.973}$Ir$_{0.027}$)$_2$As$_2$}
\def\mathbi#1{\textbf{\em #1}}

\date{\today}

\begin{abstract}
We investigate the magnetic polarization of the Ir $5d$ dopant states in the pnictide superconductor Ba(Fe$_{1-x}$Ir$_x$)$_2$As$_2$ with $x=0.027(2)$ using Ir $L_3$ edge x-ray resonant magnetic scattering (XRMS). Despite the fact that doping partially suppresses the antiferromagnetic transition, we find that magnetic order survives around the Ir dopant sites. The Ir states are magnetically polarized with commensurate stripe-like antiferromagnetic order and long correlations lengths, $\xi_{\text{mag}}>$ 2800 and $>$850~\AA{}, in the $ab$-plane and along the $c$-axis, respectively, driven by their interaction with the Fe spins. This Ir magnetic order persists up to the N\'{e}el transition of the majority Fe spins at $T_N=74(2)$~K. At 5~K we find that magnetic order co-exists microscopically with superconductivity in Ba(Fe$_{1-x}$Ir$_x$)$_2$As$_2$. The energy dependence of the XRMS through the Ir $L_3$ edge shows a non-Lorentzian lineshape, which we explain in terms of interference between Ir resonant scattering and Fe non-resonant magnetic scattering.
\end{abstract}

% insert suggested PACS numbers in braces on next line
\pacs{74.70.Xa,75.25.-j,71.70.Ej}
% insert suggested keywords - APS authors don't need to do this
%\keywords{}

%\maketitle must follow title, authors, abstract, \pacs, and \keywords
\maketitle

%\section{Introduction}
The discovery of superconductivity in the iron-pnictides has ignited intense interest in the interplay between magnetism and superconductivity in these compounds.\cite{Lumsden2010,Johnston2010,Canfield2010,Stewart2011} The 122 family of pnictides, with formula $A$Fe$_2$As$_2$ where $A=$ Ba, Ca, or Sr, is paramagnetic with a tetragonal crystal structural at room temperature before undergoing structural and magnetic transitions into an orthorhombic, antiferromagnetically ordered ground state.\cite{Huang2008,Goldman2008,Zhao2008,Kim2011} Various dopants can be substituted into any one of the three different atomic sites,\cite{Ni2011review} and act to reduce the structural and magnetic transition temperatures. Doping $A$Fe$_2$As$_2$ with K,\cite{Rotter2008Kdoping} Co,\cite{Sefat2008,Ni2008Co} Ni,\cite{Li2009,Canfield2009} Rh,\cite{Ni2009RhPd,Han2009} P,\cite{Kasahara2010} Pd,\cite{Ni2009RhPd,Han2009} Ir,\cite{Han2009} Pt,\cite{Saha2010} or Ru\cite{Sharma2010,Thaler2010,Kim2011Ru} induces superconductivity, although Cr,\cite{Marty2011} Mn,\cite{Liu2010,Thaler2011} and Mo,\cite{Sefat2012} do not. Due to the qualitatively similar effects of many different dopants, some studies have suggested that the role of the dopants in destabilizing magnetism and inducing superconductivity is simply to act as a scattering center.\cite{Wadati2010,Vavilov2011,Dhaka2011} If dopant atoms indeed act as strong scatters, this raises the question of whether the properties of pnictides are modified around the dopant states. For example, is the local value of the magnetic order parameter suppressed to zero at the dopant sites, while remaining finite globally?

% summary bit
Here we exploit the Ir $L_3$ edge resonance to isolate the magnetic behavior of the Ir $5d$ dopant states in \sample{}.  We demonstrate that the Ir states are magnetically polarized at low temperatures, with stripe-like commensurate magnetic order and long correlations lengths $>2800$~\AA{} in the $ab$-plane and $>850$~\AA{} along the $c$-axis. This Ir magnetic ordering disappears above $T_N$ and is consistent with the Ir $5d$ states being polarized via their interaction with the Fe spins. The Ir magnetic order also co-exists microscopically with superconductivity at 5~K with no evidence for phase separation. The energy dependence of the XRMS through the Ir $L_3$ resonance shows a distinct non-Lorentzian lineshape consistent with interference between resonant magnetic scattering and non-resonant magnetic scattering.

\begin{figure}
\includegraphics{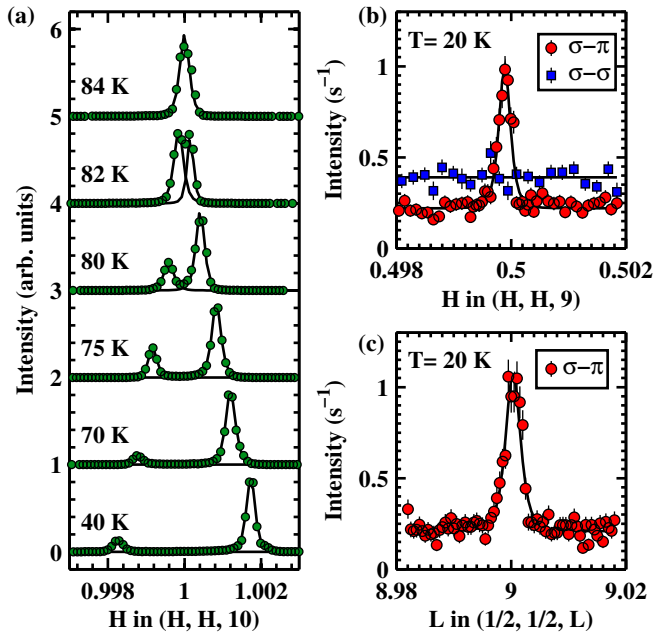} %
\caption{(color online). (a) $(H, H, 10)_T$ scans through the $(1, 1, 10)_T$ Bragg peak at several temperatures, showing the peak splitting that occurs due to the tetragonal-orthorhombic structural transition. (b) $(H, H, 9)_T$ scans through the magnetic peak at $(\frac{1}{2}, \frac{1}{2}, 9)_T$ in $\sigma-\pi$ geometry (red~$\bullet$) and $\sigma-\sigma$ geometry (blue~$\filledsquare$). (c) $L$ scan through the magnetic peak at $(\frac{1}{2}, \frac{1}{2}, 9)_T$ in $\sigma-\pi$ geometry. (b) and (c) were measured at $T= 20$~K with $E= 11.220$~keV x-rays.}
\label{fig1}
\end{figure}

%\section{Experimental methods}
Single crystals of Ba(Fe$_{1-x}$Ir$_x$)$_2$As$_2$ were prepared using the self-flux solution growth method.\cite{Ni2008Co,Canfield2009,Ni2009RhPd,Thaler2010} Wavelength-dispersive spectroscopy was employed to measure the Ir concentration at several points on several pieces from the batch, giving an Ir concentration of $x=0.027(2)$. Further attesting to the high sample quality, the crystalline mosaic was found to be 0.01$^{\circ}$ full width at half maximum (FWHM). The sample was shown to be superconducting with resistivity and magnetic susceptibility measurements.

XRMS experiments were performed at the 6ID-B beamline at the Advanced Photon Source (APS), following initial measurements on X22C at the National Synchrotron Light Source. The measurements at the APS were performed on a sample with a cleaved $c$-axis crystalline face $1 \times 2$~mm$^2$ in area with the $(H, H, L)$ plane parallel to the vertical scattering plane. The incident x-ray beam was 1.0~mm horizontal $\times$ 0.2~mm vertical in size and horizontally ($\sigma$) polarized. The sample was mounted on a Cu sample holder and surrounded with helium exchange gas. We checked for possible x-ray beam heating effects by comparing scans with and without attenuating the beam and we estimate that these effects are $<2$~K. A graphite crystal was used as polarization analyzer before the detector to distinguish $\sigma$-$\sigma$ charge scattering or fluorescence from $\sigma$-$\pi$ dipole magnetic scattering, which rotates the polarization of the incident beam. At energies around the Ir $L_3$ edge at 11.2~keV the graphite (0, 0, 10) reflection was employed while at energies around the $L_2$ edge, at 12.83~keV, the (0, 0, 8) reflection was chosen.

%% structural stuff
BaFe$_2$As$_2$ crystallizes in the tetragonal ThCr$_2$Si$_2$-type structure at room temperature with space group $I4/mmm $ (N$^\mathrm{o}$ 139) and $a=b=3.96$ and $c=13.02$~\AA{}.\cite{Rotter2008struct} Below the structural transition temperature, $T_S$, an orthorhombic distortion sets in and the crystal assumes the $Fmmm$ (N$^\mathrm{o}$ 69) space group with $a=5.61$, $b=5.57$ and $c=12.95$~\AA{} at 20~K.\cite{Rotter2008struct} In this paper we will predominately use the tetragonal notation, and where necessary use $(H, K, L)_T$ and $(H, K, L)_O$ to distinguish tetragonal and orthorhombic notations, respectively. Figure~\ref{fig1}(a) plots $(H, H, 10)_T$ scans through the $(1, 1, 10)_T$ Bragg peak on \sample{}. Below $T_S= 83$~K two split peaks are observed: the peak at lower $H$ corresponds to $(2, 0, 10)_O$; while the peak at higher $H$ corresponds to $(0, 2, 10)_O$. Below $T_S$ we use the lower $H$ peak to define the $(1, 1, 10)_T$ peak in the orientation-matrix. In this way $(\frac{1}{2}, \frac{1}{2}, L)_T$ corresponds to $(1, 0, L)_O$.

%\section{Results}
% Describe presence of magnetic peak
In order to study the Ir $5d$ dopant states, we tuned the incident x-ray energy to the Ir $L_3$ edge corresponding to exciting a $2p_{3/2}$ core electron into the $5d$ valence band. Figure~\ref{fig1}(b) plots $(H, H, 9)_T$ scans through the $(\frac{1}{2}, \frac{1}{2}, 9)_T$ position at $T=20$~K. A clear peak is present in $\sigma$-$\pi$ geometry and absent in $\sigma$-$\sigma$. This demonstrates that this peak is not associated with charge scattering such as might arise from a structural distortion. In principle $\sigma$-$\pi$ scattering could arise from either spin\cite{Hill1996} or orbital ordering,\cite{Wilkins2003,Staub2005}. However, as we shall demonstrate, this peak has all the characteristics of magnetic order, and occurs at the same wave vector as is typically observed for magnetic order in the 122 pnictides.\cite{Rotter2008struct,Lumsden2010,Johnston2010,Canfield2010,Stewart2011} Thus we conclude that it results from a magnetic polarization of the Ir $5d$ states. It should also be noted that since Ir has strong spin-orbit coupling the spin and orbital degrees of freedom are mixed. Therefore, this magnetic peak is also likely to have some partial orbital character. 

The $(H, H, 9)_T$ scan in Fig.~\ref{fig1}(b) and the $L$ scan in Fig.~\ref{fig1}(c) were fit with a Lorentzian-squared lineshape, shown as the black lines. The widths of the $(\frac{1}{2}, \frac{1}{2}, 9)_T$ peak along $H$ and $L$ are similar to that of the charge Bragg peaks, suggesting that the magnetic polarization of the Ir site is well-correlated, presumably due to coupling via the Fe spins. These widths determine lower limits on the correlation length defined as $\xi = 2a_{\text{eff}}/ (2\pi w)$, where $a_{\text{eff}}$ is the effective lattice parameter in the relevant direction and $w$ is the peak FWHM in reciprocal lattice units (r.l.u.). We find $\xi_{\text{mag}}>$ 2800 and 850~\AA{} in the $ab$-plane and along the $c$-axis, respectively. Thus  finite magnetic order survives on the Ir atom and on the neighboring Fe sites which act to polarize the Ir site. This excludes the possibility that the magnetic order parameter is locally reduced to zero around the dopant sites, as might be expected if the Ir is strongly perturbing the Fe magnetic lattice. The Ir $5d$ states may be polarized by either the local field from the Fe neighbors or via other indirect interations between the Ir and Fe states. It is difficult to distinguish these scenarios, and in this itinerant system with extended Ir $5d$ states, both effects are likely to be at work.

It is also noteworthy that the magnetic ordering vector is commensurate with the lattice, at least to within our $\mathbi{Q}$ resolution of 0.0002 r.l.u.\ in the $[1 1 0]_T$ direction. Incommensurability in $A$(Fe$_{1-x}$M$_x$)$_2$As$_2$ was suggested on the basis of local probe measurements \cite{Laplace2009,Bonville2010,Blachowski2011} and subsequently by direct observation with neutron scattering.\cite{Pratt2011}  For low doping $x<0.047$, Ba(Fe$_{1-x}$Co$_x$)$_2$As$_2$ shows commensurate magnetic order whereas incommensurate magnetic order appears for $0.056<x<0.06$ before $T_N$ is completely suppressed.\cite{Pratt2011} Whether the magnetic order in  Ba(Fe$_{1-x}$Ir$_x$)$_2$As$_2$ becomes incommensurate at higher dopings will be an interesting topic for future studies.

% resonance behavior
\begin{figure}
\includegraphics{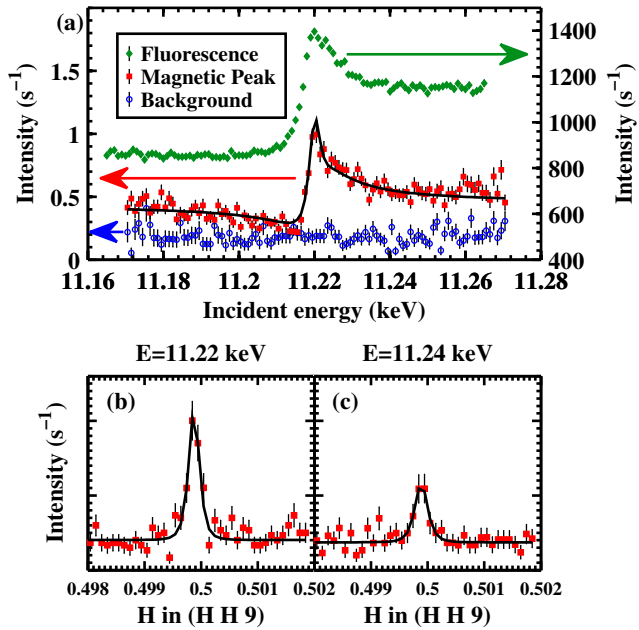} %
\caption{(color online). The Ir resonance. (a) Energy scans through the Ir $L_3$ edge plotting fluorescence yield (green~$\filleddiamond$), magnetic peak intensity at fixed $\mathbi{Q}= (\frac{1}{2}, \frac{1}{2}, 9)_T$ (red~$\filledsquare$) and the background scattering close to the magnetic peak (blue~$\bullet$). All data were measured at $T= 20$~K. (b-c) $(H, H, 9)_T$ scans through the $(\frac{1}{2}, \frac{1}{2}, 9)_T$ magnetic peak with x-ray energies of (b) $E= 11.220$~keV and (c) $E= 11.240$~keV.}
\label{fig2}
\end{figure}

In Fig.~\ref{fig2} we examine the resonant behavior of the peak. As a function of increasing incident energy, Fig.~\ref{fig2}(a) shows the XRMS intensity at $(\frac{1}{2}, \frac{1}{2}, 9)_T$ drops slightly below the Ir absorption edge, then increases sharply through the edge, before dropping off slowly above the edge. To display the position of the absorption edge, we also plot the fluorescence yield signal obtained under the same experimental conditions. Similar resonant energy dependence of the scattered signal have been observed before at $L$ edge XRMS, for example in heavy rare earth elements.\cite{Kim2005,Gibbs1991} For these elements the lineshape has been interpreted in terms of interference between resonant and non-resonant scattering from the same element. Here we propose that interference between Ir resonant scattering with non-resonant magnetic scattering, predominately from the majority Fe atoms, is the cause. Although the non-resonant scattering is very weak, it has been observed in  previous experiments on BaFe$_2$As$_2$ at the same beamline,\cite{Kim2010} and it is of the right order of magnitude for this explanation to hold. To test whether the proposed interference can account for this resonance, we examine the magnetic intensity $|f(E)|^2$, as a function of incident energy $E$, that results from the interference between a Lorentzian resonance and an energy-independent non-resonant term\cite{Kim2005,Gibbs1991}

\begin{eqnarray}
|f(E)|^2 & \propto & \Bigg| - \frac{\Gamma}{4} \left(\frac{m}{\hbar^2}\right) \frac{E_0^2 (F_{11}-F_{1-1})}{ (E_0 - E)^2 +\Gamma^2/4 } M_{\text{res}} \nonumber\\
                          & + & i \Bigg[  \frac{1}{2} \left(\frac{m}{\hbar^2}\right) \frac{E_0^2 (F_{11}-F_{1-1} )(E_0 -E) }{(E_0 - E)^2 + \Gamma^2/4} M_{\text{res}} \nonumber\\
                          & -  & \frac{\hbar \omega}{mc^2}  M_{\text{nonres}} \Bigg] \Bigg| ^2  .
\label{reseqn}
\end{eqnarray}

Here $E_0$ is the resonant energy and $\Gamma$ the FWHM of the resonance resulting from the core-hole lifetime. $M_{\text{res}}$ and $M_{\text{nonres}}$ are the polarization matrices for resonant and non-resonant scattering, respectively, which we approximate as constant over this energy range. Absorption effects of $<4$\% are neglected.\cite{Hill1996} $E_0^2 (F_{11}-F_{1-1})$ is the dipole resonant scattering amplitude.

The lineshape obtained by least-squares fitting to Eq.~\eqref{reseqn} is plotted as the black line in Fig.~\ref{fig2}(a), where a constant offset of 0.18~s$^{-1}$ has been added to account for the background. The resulting fit provides a good description of the data. In particular the reduction of intensity from $E=11.170-11.213$~keV is difficult to explain without invoking the proposed interference.

A value of $E_0=11.220$~keV was obtained for the resonance energy in the XRMS, corresponding to the observed white line of the fluorescence. This is different from other insulating iridium-based compounds such as Sr$_2$IrO$_4$,\cite{Kim2009} Na$_{2}$IrO$_{3}$,\cite{Liu2011} and Sr$_3$Ir$_2$O$_7$,\cite{Boseggia2012} where the peak in the XRMS occurs on the rising edge of the Ir $L_3$ fluorescence. The resonance width was also somewhat different with $\Gamma=5.2(4)$~eV compared to values of around 8~eV in insulating iridates,\cite{Boseggia2012} these differences may reflect differences in the Ir valence or in the crystal field environment or both.

We also searched for XRMS at the Ir $L_2$ edge, but we were not able to find any signal, which would be consistent with the $L_2$ resonance intensity being too weak for us to measure.

\begin{figure}
\includegraphics{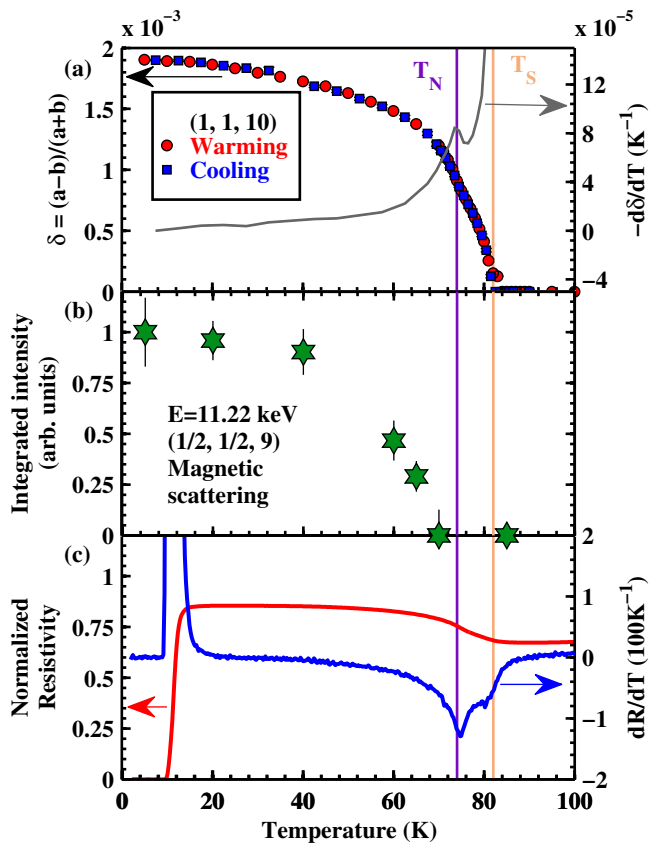} %
\caption{(color online). The temperature dependence of the magnetic and structural order parameters. (a) The magnitude of the structural distortion $\delta= (a-b)/(a+b)$ (left axis), red~$\bullet$ denote points obtained while warming; blue~$\filledsquare$ denote points obtained while cooling. The right axis plots the derivative $d\delta/dT$ in the $\delta$ values taken while warming, which exhibits a kink that occurs at $T_N$. (b) The 1D integrated intensity of the magnetic peak at $(\frac{1}{2}, \frac{1}{2}, 9)_T$ measured with $E=11.220$~keV x-rays. (c) Resistivity and the derivate of the resistivity, $dR/dT$ showing changes in slope around $T_S$ and $T_N$, and superconductivity with a 12~K onset temperature.}
\label{fig3}
\end{figure}

In Fig.~\ref{fig3} we examine the temperature dependence of the structural and magnetic ordering in \sample{}. Figure~\ref{fig3}(a) plots the magnitude of the orthorhombic distortion $\delta=(a-b)/(a+b)$, which becomes finite below $T_S=82(1)$~K in a second-order phase transition. This is substantially depressed from undoped BaFe$_2$As$_2$ with $T_S=134$~K.\cite{Rotter2008struct} On the right axis the derivative $-d\delta/dT$ is plotted to highlight a kink in $\delta$ at 74(2)~K. X-ray scattering experiments have revealed strong magneto-elastic coupling in Ba(Fe$_{1-x}$Co$_x$)$_2$As$_2$, which allows us to associate the peak in $d\delta/dT$ with the N\'{e}el transition of the Fe spins at $T_N$.\cite{Nandi2010,Kim2011} 

Going further, the Ir $L_3$ edge XRMS signal plotted in Fig.~\ref{fig3}(b), shows the magnetic polarization of the Ir $5d$ states. The ordering sets in at 70(5)~K consistent with the $T_N$ for which the Fe atoms order. Up to 65~K, the highest temperature at which magnetic order was observed, the Ir atoms remain well correlated with $\xi_{\text{mag}}^{\text{in-plane}} \gtrsim 2800$~\AA{}. The resistivity measurement in Fig.~\ref{fig3}(c) shows changes in slope at $T_S$ and $T_N$. Such changes in resistivity have been correlated with $T_S$ and $T_N$ previously in Ba(Fe$_{1-x}$Co$_x$)$_2$As$_2$.\cite{Pratt2009} The resistivity measurements also show the onset of superconductivity at 12~K. Thus in the XRMS measurements at 5~K we see Ir magnetic order coexisting with superconductivity with no measurable change in the correlation length ($\xi_{\text{mag}}^{\text{in-plane}} \gtrsim 2800$~\AA{}) above and below $T_c$. Given that the majority of Fe atoms are known to order from neutron scattering with an ordered moment of 0.60(5)~$\mathrm{\mu_B}$ \footnote{M.G. Kim \emph{et al.} unpublished} and the present experiment which demonstrates that magnetism can survive around the Ir atoms, this suggests a completely magnetic sample, such that superconductivity coexists microscopically with magnetism. This is consistent with local probe measurements of transition metal doped 122 pnictides,\cite{Julien2009,Laplace2009,Park2009,Wang2011} but in contrast to studies of Ba$_{1-x}$K$_x$Fe$-2$As$_2$.\cite{Fukazawa2009,Goko2009,Park2009}

These results establish one point in the Ba(Fe$_{1-x}$Ir$_x$)$_2$As$_2$ phase diagram with $x=0.027(2)$, $T_S=134(1)$~K, $T_N=74(2)$~K and superconductivity below 12~K. In studies of Ba(Fe$_{1-x}M_x$)$_2$As$_2$ single crystals made in the same way, $T_S$ and $T_N$ are reduced to zero by $x\gtrsim0.05$ for $M$=Co and Rh.\cite{Ni2008Co,Ni2009RhPd,Canfield2009,Ni2011review} Based on this single doping, Ir reduces $T_S$ and $T_N$ at an approximately comparable rate. Ca(Fe$_{1-x}$Ir$_x$)$_2$As$_2$ also requires similar levels of doping to suppress magnetic order or to induce superconductivity.\cite{Qi2011} In measurements of powder samples of Sr(Fe$_{1-x}$Ir$_x$)$_2$As$_2$, Han \emph{et al.}\ [Ref. \onlinecite{Han2009}] reported that far higher nominal doping values of $x\gtrsim0.2$ are required to suppress magnetic or to induce superconductivity which suggests that either Sr(Fe$_{1-x}$Ir$_x$)$_2$As$_2$ is much less sensitive to doping than Ba(Fe$_{1-x}$Ir$_x$)$_2$As$_2$ or that there is a large difference between nominal and actual doping levels in Sr(Fe$_{1-x}$Ir$_x$)$_2$As$_2$.

%\section{Conclusions}
To conclude, we have measured the magnetic polarization of the Ir $5d$ dopant states in \sample{}, which undergoes a structural phase transition at $T_S=82(1)$~K and a N\'{e}el ordering of the majority Fe spins at $T_N=74(2)$~K. Despite the fact that dopant atoms partially suppress $T_N$, we show that magnetism survives locally around the Ir sites and coexists microscopically with superconductivity at 5~K. The Ir $5d$ states are magnetically polarized with commensurate magnetic order and long correlation lengths $\xi_{\text{mag}}>$ 2800 and 850~\AA{} in the $ab$-plane and along the $c$-axis respectively, demonstrating that the Ir states are coupled via the Fe magnetism. This ordering sets in at 70(5)~K, consistent with $T_N$. The XRMS intensity as a function of x-ray energy through the Ir $L_3$ edge shows a non-Lorentzian lineshape, which we explain in terms of interference between Ir resonant scattering and Fe non-resonant magnetic scattering.

%\begin{acknowledgments}
The work at Brookhaven is supported in part under contract No.\ DEAC02-98CH10886 (JPH) and in part by the Center for Emergent Superconductivity (MPMD), an Energy Frontier Research Center funded by the US Department of Energy (DOE), Office of Basic Energy Sciences.  Work at the Ames Laboratory was supported by the Division of Materials Sciences and Engineering, Office of Basic Energy Sciences, US DOE and is operated by Iowa State University under Contract No.\ DE-AC02-07CH11358. Use of the Advanced Photon Source, an Office of Science User Facility operated for the US DOE Office of Science by Argonne National Laboratory, was supported by the US DOE under Contract No.\ DE-AC02-06CH11357. Preliminary experiments were performed at the X22C beamline at the National Synchrotron Light Source, Brookhaven National Laboratory, which is supported by the US DOE under Contract No. DE-AC02-98CH10886.
%\end{acknowledgments}

%


\begin{thebibliography}{51}%
\makeatletter
\providecommand \@ifxundefined [1]{%
 \@ifx{#1\undefined}
}%
\providecommand \@ifnum [1]{%
 \ifnum #1\expandafter \@firstoftwo
 \else \expandafter \@secondoftwo
 \fi
}%
\providecommand \@ifx [1]{%
 \ifx #1\expandafter \@firstoftwo
 \else \expandafter \@secondoftwo
 \fi
}%
\providecommand \natexlab [1]{#1}%
\providecommand \enquote  [1]{``#1''}%
\providecommand \bibnamefont  [1]{#1}%
\providecommand \bibfnamefont [1]{#1}%
\providecommand \citenamefont [1]{#1}%
\providecommand \href@noop [0]{\@secondoftwo}%
\providecommand \href [0]{\begingroup \@sanitize@url \@href}%
\providecommand \@href[1]{\@@startlink{#1}\@@href}%
\providecommand \@@href[1]{\endgroup#1\@@endlink}%
\providecommand \@sanitize@url [0]{\catcode `\\12\catcode `\$12\catcode
  `\&12\catcode `\#12\catcode `\^12\catcode `\_12\catcode `\%12\relax}%
\providecommand \@@startlink[1]{}%
\providecommand \@@endlink[0]{}%
\providecommand \url  [0]{\begingroup\@sanitize@url \@url }%
\providecommand \@url [1]{\endgroup\@href {#1}{\urlprefix }}%
\providecommand \urlprefix  [0]{URL }%
\providecommand \Eprint [0]{\href }%
\providecommand \doibase [0]{http://dx.doi.org/}%
\providecommand \selectlanguage [0]{\@gobble}%
\providecommand \bibinfo  [0]{\@secondoftwo}%
\providecommand \bibfield  [0]{\@secondoftwo}%
\providecommand \translation [1]{[#1]}%
\providecommand \BibitemOpen [0]{}%
\providecommand \bibitemStop [0]{}%
\providecommand \bibitemNoStop [0]{.\EOS\space}%
\providecommand \EOS [0]{\spacefactor3000\relax}%
\providecommand \BibitemShut  [1]{\csname bibitem#1\endcsname}%
\let\auto@bib@innerbib\@empty
%</preamble>
\bibitem [{\citenamefont {Lumsden}\ and\ \citenamefont
  {Christianson}(2010)}]{Lumsden2010}%
  \BibitemOpen
  \bibfield  {author} {\bibinfo {author} {\bibfnamefont {M.~D.}\ \bibnamefont
  {Lumsden}}\ and\ \bibinfo {author} {\bibfnamefont {A.~D.}\ \bibnamefont
  {Christianson}},\ }\href@noop {} {\bibfield  {journal} {\bibinfo  {journal}
  {Journal of Physics: Condensed Matter}\ }\textbf {\bibinfo {volume} {22}},\
  \bibinfo {pages} {203203} (\bibinfo {year} {2010})}\BibitemShut {NoStop}%
\bibitem [{\citenamefont {Johnston}(2010)}]{Johnston2010}%
  \BibitemOpen
  \bibfield  {author} {\bibinfo {author} {\bibfnamefont {D.~C.}\ \bibnamefont
  {Johnston}},\ }\href {\doibase 10.1080/00018732.2010.513480} {\bibfield
  {journal} {\bibinfo  {journal} {Advances in Physics}\ }\textbf {\bibinfo
  {volume} {59}},\ \bibinfo {pages} {803} (\bibinfo {year} {2010})}\BibitemShut
  {NoStop}%
\bibitem [{\citenamefont {Canfield}\ and\ \citenamefont
  {Bud'ko}(2010)}]{Canfield2010}%
  \BibitemOpen
  \bibfield  {author} {\bibinfo {author} {\bibfnamefont {P.~C.}\ \bibnamefont
  {Canfield}}\ and\ \bibinfo {author} {\bibfnamefont {S.~L.}\ \bibnamefont
  {Bud'ko}},\ }\href {\doibase 10.1146/annurev-conmatphys-070909-104041}
  {\bibfield  {journal} {\bibinfo  {journal} {Annual Review of Condensed Matter
  Physics}\ }\textbf {\bibinfo {volume} {1}},\ \bibinfo {pages} {27} (\bibinfo
  {year} {2010})}\BibitemShut {NoStop}%
\bibitem [{\citenamefont {Stewart}(2011)}]{Stewart2011}%
  \BibitemOpen
  \bibfield  {author} {\bibinfo {author} {\bibfnamefont {G.~R.}\ \bibnamefont
  {Stewart}},\ }\href {\doibase 10.1103/RevModPhys.83.1589} {\bibfield
  {journal} {\bibinfo  {journal} {Rev. Mod. Phys.}\ }\textbf {\bibinfo {volume}
  {83}},\ \bibinfo {pages} {1589} (\bibinfo {year} {2011})}\BibitemShut
  {NoStop}%
\bibitem [{\citenamefont {Huang}\ \emph {et~al.}(2008)\citenamefont {Huang},
  \citenamefont {Qiu}, \citenamefont {Bao}, \citenamefont {Green},
  \citenamefont {Lynn}, \citenamefont {Gasparovic}, \citenamefont {Wu},
  \citenamefont {Wu},\ and\ \citenamefont {Chen}}]{Huang2008}%
  \BibitemOpen
  \bibfield  {author} {\bibinfo {author} {\bibfnamefont {Q.}~\bibnamefont
  {Huang}}, \bibinfo {author} {\bibfnamefont {Y.}~\bibnamefont {Qiu}}, \bibinfo
  {author} {\bibfnamefont {W.}~\bibnamefont {Bao}}, \bibinfo {author}
  {\bibfnamefont {M.~A.}\ \bibnamefont {Green}}, \bibinfo {author}
  {\bibfnamefont {J.~W.}\ \bibnamefont {Lynn}}, \bibinfo {author}
  {\bibfnamefont {Y.~C.}\ \bibnamefont {Gasparovic}}, \bibinfo {author}
  {\bibfnamefont {T.}~\bibnamefont {Wu}}, \bibinfo {author} {\bibfnamefont
  {G.}~\bibnamefont {Wu}}, \ and\ \bibinfo {author} {\bibfnamefont {X.~H.}\
  \bibnamefont {Chen}},\ }\href {\doibase 10.1103/PhysRevLett.101.257003}
  {\bibfield  {journal} {\bibinfo  {journal} {Phys. Rev. Lett.}\ }\textbf
  {\bibinfo {volume} {101}},\ \bibinfo {pages} {257003} (\bibinfo {year}
  {2008})}\BibitemShut {NoStop}%
\bibitem [{\citenamefont {Goldman}\ \emph {et~al.}(2008)\citenamefont
  {Goldman}, \citenamefont {Argyriou}, \citenamefont {Ouladdiaf}, \citenamefont
  {Chatterji}, \citenamefont {Kreyssig}, \citenamefont {Nandi}, \citenamefont
  {Ni}, \citenamefont {Bud'ko}, \citenamefont {Canfield},\ and\ \citenamefont
  {McQueeney}}]{Goldman2008}%
  \BibitemOpen
  \bibfield  {author} {\bibinfo {author} {\bibfnamefont {A.~I.}\ \bibnamefont
  {Goldman}}, \bibinfo {author} {\bibfnamefont {D.~N.}\ \bibnamefont
  {Argyriou}}, \bibinfo {author} {\bibfnamefont {B.}~\bibnamefont {Ouladdiaf}},
  \bibinfo {author} {\bibfnamefont {T.}~\bibnamefont {Chatterji}}, \bibinfo
  {author} {\bibfnamefont {A.}~\bibnamefont {Kreyssig}}, \bibinfo {author}
  {\bibfnamefont {S.}~\bibnamefont {Nandi}}, \bibinfo {author} {\bibfnamefont
  {N.}~\bibnamefont {Ni}}, \bibinfo {author} {\bibfnamefont {S.~L.}\
  \bibnamefont {Bud'ko}}, \bibinfo {author} {\bibfnamefont {P.~C.}\
  \bibnamefont {Canfield}}, \ and\ \bibinfo {author} {\bibfnamefont {R.~J.}\
  \bibnamefont {McQueeney}},\ }\href {\doibase 10.1103/PhysRevB.78.100506}
  {\bibfield  {journal} {\bibinfo  {journal} {Phys. Rev. B}\ }\textbf {\bibinfo
  {volume} {78}},\ \bibinfo {pages} {100506} (\bibinfo {year}
  {2008})}\BibitemShut {NoStop}%
\bibitem [{\citenamefont {Zhao}\ \emph {et~al.}(2008)\citenamefont {Zhao},
  \citenamefont {Ratcliff}, \citenamefont {Lynn}, \citenamefont {Chen},
  \citenamefont {Luo}, \citenamefont {Wang}, \citenamefont {Hu},\ and\
  \citenamefont {Dai}}]{Zhao2008}%
  \BibitemOpen
  \bibfield  {author} {\bibinfo {author} {\bibfnamefont {J.}~\bibnamefont
  {Zhao}}, \bibinfo {author} {\bibfnamefont {W.}~\bibnamefont {Ratcliff}},
  \bibinfo {author} {\bibfnamefont {J.~W.}\ \bibnamefont {Lynn}}, \bibinfo
  {author} {\bibfnamefont {G.~F.}\ \bibnamefont {Chen}}, \bibinfo {author}
  {\bibfnamefont {J.~L.}\ \bibnamefont {Luo}}, \bibinfo {author} {\bibfnamefont
  {N.~L.}\ \bibnamefont {Wang}}, \bibinfo {author} {\bibfnamefont
  {J.}~\bibnamefont {Hu}}, \ and\ \bibinfo {author} {\bibfnamefont
  {P.}~\bibnamefont {Dai}},\ }\href {\doibase 10.1103/PhysRevB.78.140504}
  {\bibfield  {journal} {\bibinfo  {journal} {Phys. Rev. B}\ }\textbf {\bibinfo
  {volume} {78}},\ \bibinfo {pages} {140504} (\bibinfo {year}
  {2008})}\BibitemShut {NoStop}%
\bibitem [{\citenamefont {Kim}\ \emph {et~al.}(2011{\natexlab{a}})\citenamefont
  {Kim}, \citenamefont {Fernandes}, \citenamefont {Kreyssig}, \citenamefont
  {Kim}, \citenamefont {Thaler}, \citenamefont {Bud'ko}, \citenamefont
  {Canfield}, \citenamefont {McQueeney}, \citenamefont {Schmalian},\ and\
  \citenamefont {Goldman}}]{Kim2011}%
  \BibitemOpen
  \bibfield  {author} {\bibinfo {author} {\bibfnamefont {M.~G.}\ \bibnamefont
  {Kim}}, \bibinfo {author} {\bibfnamefont {R.~M.}\ \bibnamefont {Fernandes}},
  \bibinfo {author} {\bibfnamefont {A.}~\bibnamefont {Kreyssig}}, \bibinfo
  {author} {\bibfnamefont {J.~W.}\ \bibnamefont {Kim}}, \bibinfo {author}
  {\bibfnamefont {A.}~\bibnamefont {Thaler}}, \bibinfo {author} {\bibfnamefont
  {S.~L.}\ \bibnamefont {Bud'ko}}, \bibinfo {author} {\bibfnamefont {P.~C.}\
  \bibnamefont {Canfield}}, \bibinfo {author} {\bibfnamefont {R.~J.}\
  \bibnamefont {McQueeney}}, \bibinfo {author} {\bibfnamefont {J.}~\bibnamefont
  {Schmalian}}, \ and\ \bibinfo {author} {\bibfnamefont {A.~I.}\ \bibnamefont
  {Goldman}},\ }\href {\doibase 10.1103/PhysRevB.83.134522} {\bibfield
  {journal} {\bibinfo  {journal} {Phys. Rev. B}\ }\textbf {\bibinfo {volume}
  {83}},\ \bibinfo {pages} {134522} (\bibinfo {year}
  {2011}{\natexlab{a}})}\BibitemShut {NoStop}%
\bibitem [{\citenamefont {Ni}\ and\ \citenamefont
  {Bud'ko}(2011)}]{Ni2011review}%
  \BibitemOpen
  \bibfield  {author} {\bibinfo {author} {\bibfnamefont {N.}~\bibnamefont
  {Ni}}\ and\ \bibinfo {author} {\bibfnamefont {S.~L.}\ \bibnamefont
  {Bud'ko}},\ }\href {\doibase 10.1557/mrs.2011.178} {\bibfield  {journal}
  {\bibinfo  {journal} {MRS Bulletin}\ }\textbf {\bibinfo {volume} {36}},\
  \bibinfo {pages} {620} (\bibinfo {year} {2011})}\BibitemShut {NoStop}%
\bibitem [{\citenamefont {Rotter}\ \emph
  {et~al.}(2008{\natexlab{a}})\citenamefont {Rotter}, \citenamefont {Tegel},\
  and\ \citenamefont {Johrendt}}]{Rotter2008Kdoping}%
  \BibitemOpen
  \bibfield  {author} {\bibinfo {author} {\bibfnamefont {M.}~\bibnamefont
  {Rotter}}, \bibinfo {author} {\bibfnamefont {M.}~\bibnamefont {Tegel}}, \
  and\ \bibinfo {author} {\bibfnamefont {D.}~\bibnamefont {Johrendt}},\ }\href
  {\doibase 10.1103/PhysRevLett.101.107006} {\bibfield  {journal} {\bibinfo
  {journal} {Phys. Rev. Lett.}\ }\textbf {\bibinfo {volume} {101}},\ \bibinfo
  {pages} {107006} (\bibinfo {year} {2008}{\natexlab{a}})}\BibitemShut
  {NoStop}%
\bibitem [{\citenamefont {Sefat}\ \emph {et~al.}(2008)\citenamefont {Sefat},
  \citenamefont {Jin}, \citenamefont {McGuire}, \citenamefont {Sales},
  \citenamefont {Singh},\ and\ \citenamefont {Mandrus}}]{Sefat2008}%
  \BibitemOpen
  \bibfield  {author} {\bibinfo {author} {\bibfnamefont {A.~S.}\ \bibnamefont
  {Sefat}}, \bibinfo {author} {\bibfnamefont {R.}~\bibnamefont {Jin}}, \bibinfo
  {author} {\bibfnamefont {M.~A.}\ \bibnamefont {McGuire}}, \bibinfo {author}
  {\bibfnamefont {B.~C.}\ \bibnamefont {Sales}}, \bibinfo {author}
  {\bibfnamefont {D.~J.}\ \bibnamefont {Singh}}, \ and\ \bibinfo {author}
  {\bibfnamefont {D.}~\bibnamefont {Mandrus}},\ }\href {\doibase
  10.1103/PhysRevLett.101.117004} {\bibfield  {journal} {\bibinfo  {journal}
  {Phys. Rev. Lett.}\ }\textbf {\bibinfo {volume} {101}},\ \bibinfo {pages}
  {117004} (\bibinfo {year} {2008})}\BibitemShut {NoStop}%
\bibitem [{\citenamefont {Ni}\ \emph {et~al.}(2008)\citenamefont {Ni},
  \citenamefont {Tillman}, \citenamefont {Yan}, \citenamefont {Kracher},
  \citenamefont {Hannahs}, \citenamefont {Bud'ko},\ and\ \citenamefont
  {Canfield}}]{Ni2008Co}%
  \BibitemOpen
  \bibfield  {author} {\bibinfo {author} {\bibfnamefont {N.}~\bibnamefont
  {Ni}}, \bibinfo {author} {\bibfnamefont {M.~E.}\ \bibnamefont {Tillman}},
  \bibinfo {author} {\bibfnamefont {J.-Q.}\ \bibnamefont {Yan}}, \bibinfo
  {author} {\bibfnamefont {A.}~\bibnamefont {Kracher}}, \bibinfo {author}
  {\bibfnamefont {S.~T.}\ \bibnamefont {Hannahs}}, \bibinfo {author}
  {\bibfnamefont {S.~L.}\ \bibnamefont {Bud'ko}}, \ and\ \bibinfo {author}
  {\bibfnamefont {P.~C.}\ \bibnamefont {Canfield}},\ }\href {\doibase
  10.1103/PhysRevB.78.214515} {\bibfield  {journal} {\bibinfo  {journal} {Phys.
  Rev. B}\ }\textbf {\bibinfo {volume} {78}},\ \bibinfo {pages} {214515}
  (\bibinfo {year} {2008})}\BibitemShut {NoStop}%
\bibitem [{\citenamefont {Li}\ \emph {et~al.}(2009)\citenamefont {Li},
  \citenamefont {Luo}, \citenamefont {Wang}, \citenamefont {Chen},
  \citenamefont {Ren}, \citenamefont {Tao}, \citenamefont {Li}, \citenamefont
  {Lin}, \citenamefont {He}, \citenamefont {Zhu}, \citenamefont {Cao},\ and\
  \citenamefont {Xu}}]{Li2009}%
  \BibitemOpen
  \bibfield  {author} {\bibinfo {author} {\bibfnamefont {L.~J.}\ \bibnamefont
  {Li}}, \bibinfo {author} {\bibfnamefont {Y.~K.}\ \bibnamefont {Luo}},
  \bibinfo {author} {\bibfnamefont {Q.~B.}\ \bibnamefont {Wang}}, \bibinfo
  {author} {\bibfnamefont {H.}~\bibnamefont {Chen}}, \bibinfo {author}
  {\bibfnamefont {Z.}~\bibnamefont {Ren}}, \bibinfo {author} {\bibfnamefont
  {Q.}~\bibnamefont {Tao}}, \bibinfo {author} {\bibfnamefont {Y.~K.}\
  \bibnamefont {Li}}, \bibinfo {author} {\bibfnamefont {X.}~\bibnamefont
  {Lin}}, \bibinfo {author} {\bibfnamefont {M.}~\bibnamefont {He}}, \bibinfo
  {author} {\bibfnamefont {Z.~W.}\ \bibnamefont {Zhu}}, \bibinfo {author}
  {\bibfnamefont {G.~H.}\ \bibnamefont {Cao}}, \ and\ \bibinfo {author}
  {\bibfnamefont {Z.~A.}\ \bibnamefont {Xu}},\ }\href
  {http://stacks.iop.org/1367-2630/11/i=2/a=025008} {\bibfield  {journal}
  {\bibinfo  {journal} {New Journal of Physics}\ }\textbf {\bibinfo {volume}
  {11}},\ \bibinfo {pages} {025008} (\bibinfo {year} {2009})}\BibitemShut
  {NoStop}%
\bibitem [{\citenamefont {Canfield}\ \emph {et~al.}(2009)\citenamefont
  {Canfield}, \citenamefont {Bud'ko}, \citenamefont {Ni}, \citenamefont {Yan},\
  and\ \citenamefont {Kracher}}]{Canfield2009}%
  \BibitemOpen
  \bibfield  {author} {\bibinfo {author} {\bibfnamefont {P.~C.}\ \bibnamefont
  {Canfield}}, \bibinfo {author} {\bibfnamefont {S.~L.}\ \bibnamefont
  {Bud'ko}}, \bibinfo {author} {\bibfnamefont {N.}~\bibnamefont {Ni}}, \bibinfo
  {author} {\bibfnamefont {J.~Q.}\ \bibnamefont {Yan}}, \ and\ \bibinfo
  {author} {\bibfnamefont {A.}~\bibnamefont {Kracher}},\ }\href {\doibase
  10.1103/PhysRevB.80.060501} {\bibfield  {journal} {\bibinfo  {journal} {Phys.
  Rev. B}\ }\textbf {\bibinfo {volume} {80}},\ \bibinfo {pages} {060501}
  (\bibinfo {year} {2009})}\BibitemShut {NoStop}%
\bibitem [{\citenamefont {Ni}\ \emph {et~al.}(2009)\citenamefont {Ni},
  \citenamefont {Thaler}, \citenamefont {Kracher}, \citenamefont {Yan},
  \citenamefont {Bud'ko},\ and\ \citenamefont {Canfield}}]{Ni2009RhPd}%
  \BibitemOpen
  \bibfield  {author} {\bibinfo {author} {\bibfnamefont {N.}~\bibnamefont
  {Ni}}, \bibinfo {author} {\bibfnamefont {A.}~\bibnamefont {Thaler}}, \bibinfo
  {author} {\bibfnamefont {A.}~\bibnamefont {Kracher}}, \bibinfo {author}
  {\bibfnamefont {J.~Q.}\ \bibnamefont {Yan}}, \bibinfo {author} {\bibfnamefont
  {S.~L.}\ \bibnamefont {Bud'ko}}, \ and\ \bibinfo {author} {\bibfnamefont
  {P.~C.}\ \bibnamefont {Canfield}},\ }\href {\doibase
  10.1103/PhysRevB.80.024511} {\bibfield  {journal} {\bibinfo  {journal} {Phys.
  Rev. B}\ }\textbf {\bibinfo {volume} {80}},\ \bibinfo {pages} {024511}
  (\bibinfo {year} {2009})}\BibitemShut {NoStop}%
\bibitem [{\citenamefont {Han}\ \emph {et~al.}(2009)\citenamefont {Han},
  \citenamefont {Zhu}, \citenamefont {Cheng}, \citenamefont {Mu}, \citenamefont
  {Jia}, \citenamefont {Fang}, \citenamefont {Wang}, \citenamefont {Luo},
  \citenamefont {Zeng}, \citenamefont {Shen}, \citenamefont {Shan},
  \citenamefont {Ren},\ and\ \citenamefont {Wen}}]{Han2009}%
  \BibitemOpen
  \bibfield  {author} {\bibinfo {author} {\bibfnamefont {F.}~\bibnamefont
  {Han}}, \bibinfo {author} {\bibfnamefont {X.}~\bibnamefont {Zhu}}, \bibinfo
  {author} {\bibfnamefont {P.}~\bibnamefont {Cheng}}, \bibinfo {author}
  {\bibfnamefont {G.}~\bibnamefont {Mu}}, \bibinfo {author} {\bibfnamefont
  {Y.}~\bibnamefont {Jia}}, \bibinfo {author} {\bibfnamefont {L.}~\bibnamefont
  {Fang}}, \bibinfo {author} {\bibfnamefont {Y.}~\bibnamefont {Wang}}, \bibinfo
  {author} {\bibfnamefont {H.}~\bibnamefont {Luo}}, \bibinfo {author}
  {\bibfnamefont {B.}~\bibnamefont {Zeng}}, \bibinfo {author} {\bibfnamefont
  {B.}~\bibnamefont {Shen}}, \bibinfo {author} {\bibfnamefont {L.}~\bibnamefont
  {Shan}}, \bibinfo {author} {\bibfnamefont {C.}~\bibnamefont {Ren}}, \ and\
  \bibinfo {author} {\bibfnamefont {H.-H.}\ \bibnamefont {Wen}},\ }\href
  {\doibase 10.1103/PhysRevB.80.024506} {\bibfield  {journal} {\bibinfo
  {journal} {Phys. Rev. B}\ }\textbf {\bibinfo {volume} {80}},\ \bibinfo
  {pages} {024506} (\bibinfo {year} {2009})}\BibitemShut {NoStop}%
\bibitem [{\citenamefont {Kasahara}\ \emph {et~al.}(2010)\citenamefont
  {Kasahara}, \citenamefont {Shibauchi}, \citenamefont {Hashimoto},
  \citenamefont {Ikada}, \citenamefont {Tonegawa}, \citenamefont {Okazaki},
  \citenamefont {Shishido}, \citenamefont {Ikeda}, \citenamefont {Takeya},
  \citenamefont {Hirata}, \citenamefont {Terashima},\ and\ \citenamefont
  {Matsuda}}]{Kasahara2010}%
  \BibitemOpen
  \bibfield  {author} {\bibinfo {author} {\bibfnamefont {S.}~\bibnamefont
  {Kasahara}}, \bibinfo {author} {\bibfnamefont {T.}~\bibnamefont {Shibauchi}},
  \bibinfo {author} {\bibfnamefont {K.}~\bibnamefont {Hashimoto}}, \bibinfo
  {author} {\bibfnamefont {K.}~\bibnamefont {Ikada}}, \bibinfo {author}
  {\bibfnamefont {S.}~\bibnamefont {Tonegawa}}, \bibinfo {author}
  {\bibfnamefont {R.}~\bibnamefont {Okazaki}}, \bibinfo {author} {\bibfnamefont
  {H.}~\bibnamefont {Shishido}}, \bibinfo {author} {\bibfnamefont
  {H.}~\bibnamefont {Ikeda}}, \bibinfo {author} {\bibfnamefont
  {H.}~\bibnamefont {Takeya}}, \bibinfo {author} {\bibfnamefont
  {K.}~\bibnamefont {Hirata}}, \bibinfo {author} {\bibfnamefont
  {T.}~\bibnamefont {Terashima}}, \ and\ \bibinfo {author} {\bibfnamefont
  {Y.}~\bibnamefont {Matsuda}},\ }\href {\doibase 10.1103/PhysRevB.81.184519}
  {\bibfield  {journal} {\bibinfo  {journal} {Phys. Rev. B}\ }\textbf {\bibinfo
  {volume} {81}},\ \bibinfo {pages} {184519} (\bibinfo {year}
  {2010})}\BibitemShut {NoStop}%
\bibitem [{\citenamefont {Saha}\ \emph {et~al.}(2010)\citenamefont {Saha},
  \citenamefont {Drye}, \citenamefont {Kirshenbaum}, \citenamefont {Butch},
  \citenamefont {Zavalij},\ and\ \citenamefont {Paglione}}]{Saha2010}%
  \BibitemOpen
  \bibfield  {author} {\bibinfo {author} {\bibfnamefont {S.~R.}\ \bibnamefont
  {Saha}}, \bibinfo {author} {\bibfnamefont {T.}~\bibnamefont {Drye}}, \bibinfo
  {author} {\bibfnamefont {K.}~\bibnamefont {Kirshenbaum}}, \bibinfo {author}
  {\bibfnamefont {N.~P.}\ \bibnamefont {Butch}}, \bibinfo {author}
  {\bibfnamefont {P.~Y.}\ \bibnamefont {Zavalij}}, \ and\ \bibinfo {author}
  {\bibfnamefont {J.}~\bibnamefont {Paglione}},\ }\href
  {http://stacks.iop.org/0953-8984/22/i=7/a=072204} {\bibfield  {journal}
  {\bibinfo  {journal} {Journal of Physics: Condensed Matter}\ }\textbf
  {\bibinfo {volume} {22}},\ \bibinfo {pages} {072204} (\bibinfo {year}
  {2010})}\BibitemShut {NoStop}%
\bibitem [{\citenamefont {Sharma}\ \emph {et~al.}(2010)\citenamefont {Sharma},
  \citenamefont {Bharathi}, \citenamefont {Chandra}, \citenamefont {Reddy},
  \citenamefont {Paulraj}, \citenamefont {Satya}, \citenamefont {Sastry},
  \citenamefont {Gupta},\ and\ \citenamefont {Sundar}}]{Sharma2010}%
  \BibitemOpen
  \bibfield  {author} {\bibinfo {author} {\bibfnamefont {S.}~\bibnamefont
  {Sharma}}, \bibinfo {author} {\bibfnamefont {A.}~\bibnamefont {Bharathi}},
  \bibinfo {author} {\bibfnamefont {S.}~\bibnamefont {Chandra}}, \bibinfo
  {author} {\bibfnamefont {V.~R.}\ \bibnamefont {Reddy}}, \bibinfo {author}
  {\bibfnamefont {S.}~\bibnamefont {Paulraj}}, \bibinfo {author} {\bibfnamefont
  {A.~T.}\ \bibnamefont {Satya}}, \bibinfo {author} {\bibfnamefont {V.~S.}\
  \bibnamefont {Sastry}}, \bibinfo {author} {\bibfnamefont {A.}~\bibnamefont
  {Gupta}}, \ and\ \bibinfo {author} {\bibfnamefont {C.~S.}\ \bibnamefont
  {Sundar}},\ }\href {\doibase 10.1103/PhysRevB.81.174512} {\bibfield
  {journal} {\bibinfo  {journal} {Phys. Rev. B}\ }\textbf {\bibinfo {volume}
  {81}},\ \bibinfo {pages} {174512} (\bibinfo {year} {2010})}\BibitemShut
  {NoStop}%
\bibitem [{\citenamefont {Thaler}\ \emph {et~al.}(2010)\citenamefont {Thaler},
  \citenamefont {Ni}, \citenamefont {Kracher}, \citenamefont {Yan},
  \citenamefont {Bud'ko},\ and\ \citenamefont {Canfield}}]{Thaler2010}%
  \BibitemOpen
  \bibfield  {author} {\bibinfo {author} {\bibfnamefont {A.}~\bibnamefont
  {Thaler}}, \bibinfo {author} {\bibfnamefont {N.}~\bibnamefont {Ni}}, \bibinfo
  {author} {\bibfnamefont {A.}~\bibnamefont {Kracher}}, \bibinfo {author}
  {\bibfnamefont {J.~Q.}\ \bibnamefont {Yan}}, \bibinfo {author} {\bibfnamefont
  {S.~L.}\ \bibnamefont {Bud'ko}}, \ and\ \bibinfo {author} {\bibfnamefont
  {P.~C.}\ \bibnamefont {Canfield}},\ }\href {\doibase
  10.1103/PhysRevB.82.014534} {\bibfield  {journal} {\bibinfo  {journal} {Phys.
  Rev. B}\ }\textbf {\bibinfo {volume} {82}},\ \bibinfo {pages} {014534}
  (\bibinfo {year} {2010})}\BibitemShut {NoStop}%
\bibitem [{\citenamefont {Kim}\ \emph {et~al.}(2011{\natexlab{b}})\citenamefont
  {Kim}, \citenamefont {Pratt}, \citenamefont {Rustan}, \citenamefont {Tian},
  \citenamefont {Zarestky}, \citenamefont {Thaler}, \citenamefont {Bud'ko},
  \citenamefont {Canfield}, \citenamefont {McQueeney}, \citenamefont
  {Kreyssig},\ and\ \citenamefont {Goldman}}]{Kim2011Ru}%
  \BibitemOpen
  \bibfield  {author} {\bibinfo {author} {\bibfnamefont {M.~G.}\ \bibnamefont
  {Kim}}, \bibinfo {author} {\bibfnamefont {D.~K.}\ \bibnamefont {Pratt}},
  \bibinfo {author} {\bibfnamefont {G.~E.}\ \bibnamefont {Rustan}}, \bibinfo
  {author} {\bibfnamefont {W.}~\bibnamefont {Tian}}, \bibinfo {author}
  {\bibfnamefont {J.~L.}\ \bibnamefont {Zarestky}}, \bibinfo {author}
  {\bibfnamefont {A.}~\bibnamefont {Thaler}}, \bibinfo {author} {\bibfnamefont
  {S.~L.}\ \bibnamefont {Bud'ko}}, \bibinfo {author} {\bibfnamefont {P.~C.}\
  \bibnamefont {Canfield}}, \bibinfo {author} {\bibfnamefont {R.~J.}\
  \bibnamefont {McQueeney}}, \bibinfo {author} {\bibfnamefont {A.}~\bibnamefont
  {Kreyssig}}, \ and\ \bibinfo {author} {\bibfnamefont {A.~I.}\ \bibnamefont
  {Goldman}},\ }\href {\doibase 10.1103/PhysRevB.83.054514} {\bibfield
  {journal} {\bibinfo  {journal} {Phys. Rev. B}\ }\textbf {\bibinfo {volume}
  {83}},\ \bibinfo {pages} {054514} (\bibinfo {year}
  {2011}{\natexlab{b}})}\BibitemShut {NoStop}%
\bibitem [{\citenamefont {Marty}\ \emph {et~al.}(2011)\citenamefont {Marty},
  \citenamefont {Christianson}, \citenamefont {Wang}, \citenamefont {Matsuda},
  \citenamefont {Cao}, \citenamefont {VanBebber}, \citenamefont {Zarestky},
  \citenamefont {Singh}, \citenamefont {Sefat},\ and\ \citenamefont
  {Lumsden}}]{Marty2011}%
  \BibitemOpen
  \bibfield  {author} {\bibinfo {author} {\bibfnamefont {K.}~\bibnamefont
  {Marty}}, \bibinfo {author} {\bibfnamefont {A.~D.}\ \bibnamefont
  {Christianson}}, \bibinfo {author} {\bibfnamefont {C.~H.}\ \bibnamefont
  {Wang}}, \bibinfo {author} {\bibfnamefont {M.}~\bibnamefont {Matsuda}},
  \bibinfo {author} {\bibfnamefont {H.}~\bibnamefont {Cao}}, \bibinfo {author}
  {\bibfnamefont {L.~H.}\ \bibnamefont {VanBebber}}, \bibinfo {author}
  {\bibfnamefont {J.~L.}\ \bibnamefont {Zarestky}}, \bibinfo {author}
  {\bibfnamefont {D.~J.}\ \bibnamefont {Singh}}, \bibinfo {author}
  {\bibfnamefont {A.~S.}\ \bibnamefont {Sefat}}, \ and\ \bibinfo {author}
  {\bibfnamefont {M.~D.}\ \bibnamefont {Lumsden}},\ }\href {\doibase
  10.1103/PhysRevB.83.060509} {\bibfield  {journal} {\bibinfo  {journal} {Phys.
  Rev. B}\ }\textbf {\bibinfo {volume} {83}},\ \bibinfo {pages} {060509}
  (\bibinfo {year} {2011})}\BibitemShut {NoStop}%
\bibitem [{\citenamefont {Liu}\ \emph {et~al.}(2010)\citenamefont {Liu},
  \citenamefont {Sun}, \citenamefont {Park},\ and\ \citenamefont
  {Lin}}]{Liu2010}%
  \BibitemOpen
  \bibfield  {author} {\bibinfo {author} {\bibfnamefont {Y.}~\bibnamefont
  {Liu}}, \bibinfo {author} {\bibfnamefont {D.}~\bibnamefont {Sun}}, \bibinfo
  {author} {\bibfnamefont {J.}~\bibnamefont {Park}}, \ and\ \bibinfo {author}
  {\bibfnamefont {C.}~\bibnamefont {Lin}},\ }\href {\doibase
  10.1016/j.physc.2009.11.024} {\bibfield  {journal} {\bibinfo  {journal}
  {Physica C}\ }\textbf {\bibinfo {volume} {470}},\ \bibinfo {pages} {S513 }
  (\bibinfo {year} {2010})}\BibitemShut {NoStop}%
\bibitem [{\citenamefont {Thaler}\ \emph {et~al.}(2011)\citenamefont {Thaler},
  \citenamefont {Hodovanets}, \citenamefont {Torikachvili}, \citenamefont
  {Ran}, \citenamefont {Kracher}, \citenamefont {Straszheim}, \citenamefont
  {Yan}, \citenamefont {Mun},\ and\ \citenamefont {Canfield}}]{Thaler2011}%
  \BibitemOpen
  \bibfield  {author} {\bibinfo {author} {\bibfnamefont {A.}~\bibnamefont
  {Thaler}}, \bibinfo {author} {\bibfnamefont {H.}~\bibnamefont {Hodovanets}},
  \bibinfo {author} {\bibfnamefont {M.~S.}\ \bibnamefont {Torikachvili}},
  \bibinfo {author} {\bibfnamefont {S.}~\bibnamefont {Ran}}, \bibinfo {author}
  {\bibfnamefont {A.}~\bibnamefont {Kracher}}, \bibinfo {author} {\bibfnamefont
  {W.}~\bibnamefont {Straszheim}}, \bibinfo {author} {\bibfnamefont {J.~Q.}\
  \bibnamefont {Yan}}, \bibinfo {author} {\bibfnamefont {E.}~\bibnamefont
  {Mun}}, \ and\ \bibinfo {author} {\bibfnamefont {P.~C.}\ \bibnamefont
  {Canfield}},\ }\href {\doibase 10.1103/PhysRevB.84.144528} {\bibfield
  {journal} {\bibinfo  {journal} {Phys. Rev. B}\ }\textbf {\bibinfo {volume}
  {84}},\ \bibinfo {pages} {144528} (\bibinfo {year} {2011})}\BibitemShut
  {NoStop}%
\bibitem [{\citenamefont {Sefat}\ \emph {et~al.}(2012)\citenamefont {Sefat},
  \citenamefont {Marty}, \citenamefont {Christianson}, \citenamefont {Saparov},
  \citenamefont {McGuire}, \citenamefont {Lumsden}, \citenamefont {Tian},\ and\
  \citenamefont {Sales}}]{Sefat2012}%
  \BibitemOpen
  \bibfield  {author} {\bibinfo {author} {\bibfnamefont {A.~S.}\ \bibnamefont
  {Sefat}}, \bibinfo {author} {\bibfnamefont {K.}~\bibnamefont {Marty}},
  \bibinfo {author} {\bibfnamefont {A.~D.}\ \bibnamefont {Christianson}},
  \bibinfo {author} {\bibfnamefont {B.}~\bibnamefont {Saparov}}, \bibinfo
  {author} {\bibfnamefont {M.~A.}\ \bibnamefont {McGuire}}, \bibinfo {author}
  {\bibfnamefont {M.~D.}\ \bibnamefont {Lumsden}}, \bibinfo {author}
  {\bibfnamefont {W.}~\bibnamefont {Tian}}, \ and\ \bibinfo {author}
  {\bibfnamefont {B.~C.}\ \bibnamefont {Sales}},\ }\href {\doibase
  10.1103/PhysRevB.85.024503} {\bibfield  {journal} {\bibinfo  {journal} {Phys.
  Rev. B}\ }\textbf {\bibinfo {volume} {85}},\ \bibinfo {pages} {024503}
  (\bibinfo {year} {2012})}\BibitemShut {NoStop}%
\bibitem [{\citenamefont {Wadati}\ \emph {et~al.}(2010)\citenamefont {Wadati},
  \citenamefont {Elfimov},\ and\ \citenamefont {Sawatzky}}]{Wadati2010}%
  \BibitemOpen
  \bibfield  {author} {\bibinfo {author} {\bibfnamefont {H.}~\bibnamefont
  {Wadati}}, \bibinfo {author} {\bibfnamefont {I.}~\bibnamefont {Elfimov}}, \
  and\ \bibinfo {author} {\bibfnamefont {G.~A.}\ \bibnamefont {Sawatzky}},\
  }\href {\doibase 10.1103/PhysRevLett.105.157004} {\bibfield  {journal}
  {\bibinfo  {journal} {Phys. Rev. Lett.}\ }\textbf {\bibinfo {volume} {105}},\
  \bibinfo {pages} {157004} (\bibinfo {year} {2010})}\BibitemShut {NoStop}%
\bibitem [{\citenamefont {Vavilov}\ and\ \citenamefont
  {Chubukov}(2011)}]{Vavilov2011}%
  \BibitemOpen
  \bibfield  {author} {\bibinfo {author} {\bibfnamefont {M.~G.}\ \bibnamefont
  {Vavilov}}\ and\ \bibinfo {author} {\bibfnamefont {A.~V.}\ \bibnamefont
  {Chubukov}},\ }\href {\doibase 10.1103/PhysRevB.84.214521} {\bibfield
  {journal} {\bibinfo  {journal} {Phys. Rev. B}\ }\textbf {\bibinfo {volume}
  {84}},\ \bibinfo {pages} {214521} (\bibinfo {year} {2011})}\BibitemShut
  {NoStop}%
\bibitem [{\citenamefont {Dhaka}\ \emph {et~al.}(2011)\citenamefont {Dhaka},
  \citenamefont {Liu}, \citenamefont {Fernandes}, \citenamefont {Jiang},
  \citenamefont {Strehlow}, \citenamefont {Kondo}, \citenamefont {Thaler},
  \citenamefont {Schmalian}, \citenamefont {Bud'ko}, \citenamefont {Canfield},\
  and\ \citenamefont {Kaminski}}]{Dhaka2011}%
  \BibitemOpen
  \bibfield  {author} {\bibinfo {author} {\bibfnamefont {R.~S.}\ \bibnamefont
  {Dhaka}}, \bibinfo {author} {\bibfnamefont {C.}~\bibnamefont {Liu}}, \bibinfo
  {author} {\bibfnamefont {R.~M.}\ \bibnamefont {Fernandes}}, \bibinfo {author}
  {\bibfnamefont {R.}~\bibnamefont {Jiang}}, \bibinfo {author} {\bibfnamefont
  {C.~P.}\ \bibnamefont {Strehlow}}, \bibinfo {author} {\bibfnamefont
  {T.}~\bibnamefont {Kondo}}, \bibinfo {author} {\bibfnamefont
  {A.}~\bibnamefont {Thaler}}, \bibinfo {author} {\bibfnamefont
  {J.}~\bibnamefont {Schmalian}}, \bibinfo {author} {\bibfnamefont {S.~L.}\
  \bibnamefont {Bud'ko}}, \bibinfo {author} {\bibfnamefont {P.~C.}\
  \bibnamefont {Canfield}}, \ and\ \bibinfo {author} {\bibfnamefont
  {A.}~\bibnamefont {Kaminski}},\ }\href {\doibase
  10.1103/PhysRevLett.107.267002} {\bibfield  {journal} {\bibinfo  {journal}
  {Phys. Rev. Lett.}\ }\textbf {\bibinfo {volume} {107}},\ \bibinfo {pages}
  {267002} (\bibinfo {year} {2011})}\BibitemShut {NoStop}%
\bibitem [{\citenamefont {Rotter}\ \emph
  {et~al.}(2008{\natexlab{b}})\citenamefont {Rotter}, \citenamefont {Tegel},
  \citenamefont {Johrendt}, \citenamefont {Schellenberg}, \citenamefont
  {Hermes},\ and\ \citenamefont {P\"ottgen}}]{Rotter2008struct}%
  \BibitemOpen
  \bibfield  {author} {\bibinfo {author} {\bibfnamefont {M.}~\bibnamefont
  {Rotter}}, \bibinfo {author} {\bibfnamefont {M.}~\bibnamefont {Tegel}},
  \bibinfo {author} {\bibfnamefont {D.}~\bibnamefont {Johrendt}}, \bibinfo
  {author} {\bibfnamefont {I.}~\bibnamefont {Schellenberg}}, \bibinfo {author}
  {\bibfnamefont {W.}~\bibnamefont {Hermes}}, \ and\ \bibinfo {author}
  {\bibfnamefont {R.}~\bibnamefont {P\"ottgen}},\ }\href {\doibase
  10.1103/PhysRevB.78.020503} {\bibfield  {journal} {\bibinfo  {journal} {Phys.
  Rev. B}\ }\textbf {\bibinfo {volume} {78}},\ \bibinfo {pages} {020503}
  (\bibinfo {year} {2008}{\natexlab{b}})}\BibitemShut {NoStop}%
\bibitem [{\citenamefont {Hill}\ and\ \citenamefont
  {McMorrow}(1996)}]{Hill1996}%
  \BibitemOpen
  \bibfield  {author} {\bibinfo {author} {\bibfnamefont {J.~P.}\ \bibnamefont
  {Hill}}\ and\ \bibinfo {author} {\bibfnamefont {D.~F.}\ \bibnamefont
  {McMorrow}},\ }\href {\doibase 10.1107/S0108767395012670} {\bibfield
  {journal} {\bibinfo  {journal} {Acta Crystallographica Section A}\ }\textbf
  {\bibinfo {volume} {52}},\ \bibinfo {pages} {236} (\bibinfo {year}
  {1996})}\BibitemShut {NoStop}%
\bibitem [{\citenamefont {Wilkins}\ \emph {et~al.}(2003)\citenamefont
  {Wilkins}, \citenamefont {Spencer}, \citenamefont {Hatton}, \citenamefont
  {Collins}, \citenamefont {Roper}, \citenamefont {Prabhakaran},\ and\
  \citenamefont {Boothroyd}}]{Wilkins2003}%
  \BibitemOpen
  \bibfield  {author} {\bibinfo {author} {\bibfnamefont {S.~B.}\ \bibnamefont
  {Wilkins}}, \bibinfo {author} {\bibfnamefont {P.~D.}\ \bibnamefont
  {Spencer}}, \bibinfo {author} {\bibfnamefont {P.~D.}\ \bibnamefont {Hatton}},
  \bibinfo {author} {\bibfnamefont {S.~P.}\ \bibnamefont {Collins}}, \bibinfo
  {author} {\bibfnamefont {M.~D.}\ \bibnamefont {Roper}}, \bibinfo {author}
  {\bibfnamefont {D.}~\bibnamefont {Prabhakaran}}, \ and\ \bibinfo {author}
  {\bibfnamefont {A.~T.}\ \bibnamefont {Boothroyd}},\ }\href {\doibase
  10.1103/PhysRevLett.91.167205} {\bibfield  {journal} {\bibinfo  {journal}
  {Phys. Rev. Lett.}\ }\textbf {\bibinfo {volume} {91}},\ \bibinfo {pages}
  {167205} (\bibinfo {year} {2003})}\BibitemShut {NoStop}%
\bibitem [{\citenamefont {Staub}\ \emph {et~al.}(2005)\citenamefont {Staub},
  \citenamefont {Scagnoli}, \citenamefont {Mulders}, \citenamefont {Katsumata},
  \citenamefont {Honda}, \citenamefont {Grimmer}, \citenamefont {Horisberger},\
  and\ \citenamefont {Tonnerre}}]{Staub2005}%
  \BibitemOpen
  \bibfield  {author} {\bibinfo {author} {\bibfnamefont {U.}~\bibnamefont
  {Staub}}, \bibinfo {author} {\bibfnamefont {V.}~\bibnamefont {Scagnoli}},
  \bibinfo {author} {\bibfnamefont {A.~M.}\ \bibnamefont {Mulders}}, \bibinfo
  {author} {\bibfnamefont {K.}~\bibnamefont {Katsumata}}, \bibinfo {author}
  {\bibfnamefont {Z.}~\bibnamefont {Honda}}, \bibinfo {author} {\bibfnamefont
  {H.}~\bibnamefont {Grimmer}}, \bibinfo {author} {\bibfnamefont
  {M.}~\bibnamefont {Horisberger}}, \ and\ \bibinfo {author} {\bibfnamefont
  {J.~M.}\ \bibnamefont {Tonnerre}},\ }\href {\doibase
  10.1103/PhysRevB.71.214421} {\bibfield  {journal} {\bibinfo  {journal} {Phys.
  Rev. B}\ }\textbf {\bibinfo {volume} {71}},\ \bibinfo {pages} {214421}
  (\bibinfo {year} {2005})}\BibitemShut {NoStop}%
\bibitem [{\citenamefont {Laplace}\ \emph {et~al.}(2009)\citenamefont
  {Laplace}, \citenamefont {Bobroff}, \citenamefont {Rullier-Albenque},
  \citenamefont {Colson},\ and\ \citenamefont {Forget}}]{Laplace2009}%
  \BibitemOpen
  \bibfield  {author} {\bibinfo {author} {\bibfnamefont {Y.}~\bibnamefont
  {Laplace}}, \bibinfo {author} {\bibfnamefont {J.}~\bibnamefont {Bobroff}},
  \bibinfo {author} {\bibfnamefont {F.}~\bibnamefont {Rullier-Albenque}},
  \bibinfo {author} {\bibfnamefont {D.}~\bibnamefont {Colson}}, \ and\ \bibinfo
  {author} {\bibfnamefont {A.}~\bibnamefont {Forget}},\ }\href {\doibase
  10.1103/PhysRevB.80.140501} {\bibfield  {journal} {\bibinfo  {journal} {Phys.
  Rev. B}\ }\textbf {\bibinfo {volume} {80}},\ \bibinfo {pages} {140501}
  (\bibinfo {year} {2009})}\BibitemShut {NoStop}%
\bibitem [{\citenamefont {Bonville}\ \emph {et~al.}(2010)\citenamefont
  {Bonville}, \citenamefont {Rullier-Albenque}, \citenamefont {Colson},\ and\
  \citenamefont {Forget}}]{Bonville2010}%
  \BibitemOpen
  \bibfield  {author} {\bibinfo {author} {\bibfnamefont {P.}~\bibnamefont
  {Bonville}}, \bibinfo {author} {\bibfnamefont {F.}~\bibnamefont
  {Rullier-Albenque}}, \bibinfo {author} {\bibfnamefont {D.}~\bibnamefont
  {Colson}}, \ and\ \bibinfo {author} {\bibfnamefont {A.}~\bibnamefont
  {Forget}},\ }\href@noop {} {\bibfield  {journal} {\bibinfo  {journal} {EPL
  (Europhysics Letters)}\ }\textbf {\bibinfo {volume} {89}},\ \bibinfo {pages}
  {67008} (\bibinfo {year} {2010})}\BibitemShut {NoStop}%
\bibitem [{\citenamefont {B\l{}achowski}\ \emph {et~al.}(2011)\citenamefont
  {B\l{}achowski}, \citenamefont {Ruebenbauer}, \citenamefont
  {\ifmmode~\dot{Z}\else \.{Z}\fi{}ukrowski}, \citenamefont {Rogacki},
  \citenamefont {Bukowski},\ and\ \citenamefont {Karpinski}}]{Blachowski2011}%
  \BibitemOpen
  \bibfield  {author} {\bibinfo {author} {\bibfnamefont {A.}~\bibnamefont
  {B\l{}achowski}}, \bibinfo {author} {\bibfnamefont {K.}~\bibnamefont
  {Ruebenbauer}}, \bibinfo {author} {\bibfnamefont {J.}~\bibnamefont
  {\ifmmode~\dot{Z}\else \.{Z}\fi{}ukrowski}}, \bibinfo {author} {\bibfnamefont
  {K.}~\bibnamefont {Rogacki}}, \bibinfo {author} {\bibfnamefont
  {Z.}~\bibnamefont {Bukowski}}, \ and\ \bibinfo {author} {\bibfnamefont
  {J.}~\bibnamefont {Karpinski}},\ }\href {\doibase 10.1103/PhysRevB.83.134410}
  {\bibfield  {journal} {\bibinfo  {journal} {Phys. Rev. B}\ }\textbf {\bibinfo
  {volume} {83}},\ \bibinfo {pages} {134410} (\bibinfo {year}
  {2011})}\BibitemShut {NoStop}%
\bibitem [{\citenamefont {Pratt}\ \emph {et~al.}(2011)\citenamefont {Pratt},
  \citenamefont {Kim}, \citenamefont {Kreyssig}, \citenamefont {Lee},
  \citenamefont {Tucker}, \citenamefont {Thaler}, \citenamefont {Tian},
  \citenamefont {Zarestky}, \citenamefont {Bud'ko}, \citenamefont {Canfield},
  \citenamefont {Harmon}, \citenamefont {Goldman},\ and\ \citenamefont
  {McQueeney}}]{Pratt2011}%
  \BibitemOpen
  \bibfield  {author} {\bibinfo {author} {\bibfnamefont {D.~K.}\ \bibnamefont
  {Pratt}}, \bibinfo {author} {\bibfnamefont {M.~G.}\ \bibnamefont {Kim}},
  \bibinfo {author} {\bibfnamefont {A.}~\bibnamefont {Kreyssig}}, \bibinfo
  {author} {\bibfnamefont {Y.~B.}\ \bibnamefont {Lee}}, \bibinfo {author}
  {\bibfnamefont {G.~S.}\ \bibnamefont {Tucker}}, \bibinfo {author}
  {\bibfnamefont {A.}~\bibnamefont {Thaler}}, \bibinfo {author} {\bibfnamefont
  {W.}~\bibnamefont {Tian}}, \bibinfo {author} {\bibfnamefont {J.~L.}\
  \bibnamefont {Zarestky}}, \bibinfo {author} {\bibfnamefont {S.~L.}\
  \bibnamefont {Bud'ko}}, \bibinfo {author} {\bibfnamefont {P.~C.}\
  \bibnamefont {Canfield}}, \bibinfo {author} {\bibfnamefont {B.~N.}\
  \bibnamefont {Harmon}}, \bibinfo {author} {\bibfnamefont {A.~I.}\
  \bibnamefont {Goldman}}, \ and\ \bibinfo {author} {\bibfnamefont {R.~J.}\
  \bibnamefont {McQueeney}},\ }\href {\doibase 10.1103/PhysRevLett.106.257001}
  {\bibfield  {journal} {\bibinfo  {journal} {Phys. Rev. Lett.}\ }\textbf
  {\bibinfo {volume} {106}},\ \bibinfo {pages} {257001} (\bibinfo {year}
  {2011})}\BibitemShut {NoStop}%
\bibitem [{\citenamefont {Kim}\ \emph {et~al.}(2005)\citenamefont {Kim},
  \citenamefont {Lee}, \citenamefont {Wermeille}, \citenamefont {Sieve},
  \citenamefont {Tan}, \citenamefont {Bud'ko}, \citenamefont {Law},
  \citenamefont {Canfield}, \citenamefont {Harmon},\ and\ \citenamefont
  {Goldman}}]{Kim2005}%
  \BibitemOpen
  \bibfield  {author} {\bibinfo {author} {\bibfnamefont {J.~W.}\ \bibnamefont
  {Kim}}, \bibinfo {author} {\bibfnamefont {Y.}~\bibnamefont {Lee}}, \bibinfo
  {author} {\bibfnamefont {D.}~\bibnamefont {Wermeille}}, \bibinfo {author}
  {\bibfnamefont {B.}~\bibnamefont {Sieve}}, \bibinfo {author} {\bibfnamefont
  {L.}~\bibnamefont {Tan}}, \bibinfo {author} {\bibfnamefont {S.~L.}\
  \bibnamefont {Bud'ko}}, \bibinfo {author} {\bibfnamefont {S.}~\bibnamefont
  {Law}}, \bibinfo {author} {\bibfnamefont {P.~C.}\ \bibnamefont {Canfield}},
  \bibinfo {author} {\bibfnamefont {B.~N.}\ \bibnamefont {Harmon}}, \ and\
  \bibinfo {author} {\bibfnamefont {A.~I.}\ \bibnamefont {Goldman}},\ }\href
  {\doibase 10.1103/PhysRevB.72.064403} {\bibfield  {journal} {\bibinfo
  {journal} {Phys. Rev. B}\ }\textbf {\bibinfo {volume} {72}},\ \bibinfo
  {pages} {064403} (\bibinfo {year} {2005})}\BibitemShut {NoStop}%
\bibitem [{\citenamefont {Gibbs}\ \emph {et~al.}(1991)\citenamefont {Gibbs},
  \citenamefont {Gr\"ubel}, \citenamefont {Harshman}, \citenamefont {Isaacs},
  \citenamefont {McWhan}, \citenamefont {Mills},\ and\ \citenamefont
  {Vettier}}]{Gibbs1991}%
  \BibitemOpen
  \bibfield  {author} {\bibinfo {author} {\bibfnamefont {D.}~\bibnamefont
  {Gibbs}}, \bibinfo {author} {\bibfnamefont {G.}~\bibnamefont {Gr\"ubel}},
  \bibinfo {author} {\bibfnamefont {D.~R.}\ \bibnamefont {Harshman}}, \bibinfo
  {author} {\bibfnamefont {E.~D.}\ \bibnamefont {Isaacs}}, \bibinfo {author}
  {\bibfnamefont {D.~B.}\ \bibnamefont {McWhan}}, \bibinfo {author}
  {\bibfnamefont {D.}~\bibnamefont {Mills}}, \ and\ \bibinfo {author}
  {\bibfnamefont {C.}~\bibnamefont {Vettier}},\ }\href {\doibase
  10.1103/PhysRevB.43.5663} {\bibfield  {journal} {\bibinfo  {journal} {Phys.
  Rev. B}\ }\textbf {\bibinfo {volume} {43}},\ \bibinfo {pages} {5663}
  (\bibinfo {year} {1991})}\BibitemShut {NoStop}%
\bibitem [{\citenamefont {Kim}\ \emph {et~al.}(2010)\citenamefont {Kim},
  \citenamefont {Kreyssig}, \citenamefont {Lee}, \citenamefont {Kim},
  \citenamefont {Pratt}, \citenamefont {Thaler}, \citenamefont {Bud'ko},
  \citenamefont {Canfield}, \citenamefont {Harmon}, \citenamefont {McQueeney},\
  and\ \citenamefont {Goldman}}]{Kim2010}%
  \BibitemOpen
  \bibfield  {author} {\bibinfo {author} {\bibfnamefont {M.~G.}\ \bibnamefont
  {Kim}}, \bibinfo {author} {\bibfnamefont {A.}~\bibnamefont {Kreyssig}},
  \bibinfo {author} {\bibfnamefont {Y.~B.}\ \bibnamefont {Lee}}, \bibinfo
  {author} {\bibfnamefont {J.~W.}\ \bibnamefont {Kim}}, \bibinfo {author}
  {\bibfnamefont {D.~K.}\ \bibnamefont {Pratt}}, \bibinfo {author}
  {\bibfnamefont {A.}~\bibnamefont {Thaler}}, \bibinfo {author} {\bibfnamefont
  {S.~L.}\ \bibnamefont {Bud'ko}}, \bibinfo {author} {\bibfnamefont {P.~C.}\
  \bibnamefont {Canfield}}, \bibinfo {author} {\bibfnamefont {B.~N.}\
  \bibnamefont {Harmon}}, \bibinfo {author} {\bibfnamefont {R.~J.}\
  \bibnamefont {McQueeney}}, \ and\ \bibinfo {author} {\bibfnamefont {A.~I.}\
  \bibnamefont {Goldman}},\ }\href {\doibase 10.1103/PhysRevB.82.180412}
  {\bibfield  {journal} {\bibinfo  {journal} {Phys. Rev. B}\ }\textbf {\bibinfo
  {volume} {82}},\ \bibinfo {pages} {180412} (\bibinfo {year}
  {2010})}\BibitemShut {NoStop}%
\bibitem [{\citenamefont {Kim}\ \emph {et~al.}(2009)\citenamefont {Kim},
  \citenamefont {Ohsumi}, \citenamefont {Komesu}, \citenamefont {Sakai},
  \citenamefont {Morita}, \citenamefont {Takagi},\ and\ \citenamefont
  {Arima}}]{Kim2009}%
  \BibitemOpen
  \bibfield  {author} {\bibinfo {author} {\bibfnamefont {B.~J.}\ \bibnamefont
  {Kim}}, \bibinfo {author} {\bibfnamefont {H.}~\bibnamefont {Ohsumi}},
  \bibinfo {author} {\bibfnamefont {T.}~\bibnamefont {Komesu}}, \bibinfo
  {author} {\bibfnamefont {S.}~\bibnamefont {Sakai}}, \bibinfo {author}
  {\bibfnamefont {T.}~\bibnamefont {Morita}}, \bibinfo {author} {\bibfnamefont
  {H.}~\bibnamefont {Takagi}}, \ and\ \bibinfo {author} {\bibfnamefont
  {T.}~\bibnamefont {Arima}},\ }\href {\doibase 10.1126/science.1167106} {\
  \textbf {\bibinfo {volume} {323}},\ \bibinfo {pages} {1329} (\bibinfo {year}
  {2009})}\BibitemShut {NoStop}%
\bibitem [{\citenamefont {Liu}\ \emph {et~al.}(2011)\citenamefont {Liu},
  \citenamefont {Berlijn}, \citenamefont {Yin}, \citenamefont {Ku},
  \citenamefont {Tsvelik}, \citenamefont {Kim}, \citenamefont {Gretarsson},
  \citenamefont {Singh}, \citenamefont {Gegenwart},\ and\ \citenamefont
  {Hill}}]{Liu2011}%
  \BibitemOpen
  \bibfield  {author} {\bibinfo {author} {\bibfnamefont {X.}~\bibnamefont
  {Liu}}, \bibinfo {author} {\bibfnamefont {T.}~\bibnamefont {Berlijn}},
  \bibinfo {author} {\bibfnamefont {W.-G.}\ \bibnamefont {Yin}}, \bibinfo
  {author} {\bibfnamefont {W.}~\bibnamefont {Ku}}, \bibinfo {author}
  {\bibfnamefont {A.}~\bibnamefont {Tsvelik}}, \bibinfo {author} {\bibfnamefont
  {Y.-J.}\ \bibnamefont {Kim}}, \bibinfo {author} {\bibfnamefont
  {H.}~\bibnamefont {Gretarsson}}, \bibinfo {author} {\bibfnamefont
  {Y.}~\bibnamefont {Singh}}, \bibinfo {author} {\bibfnamefont
  {P.}~\bibnamefont {Gegenwart}}, \ and\ \bibinfo {author} {\bibfnamefont
  {J.~P.}\ \bibnamefont {Hill}},\ }\href {\doibase 10.1103/PhysRevB.83.220403}
  {\bibfield  {journal} {\bibinfo  {journal} {Phys. Rev. B}\ }\textbf {\bibinfo
  {volume} {83}},\ \bibinfo {pages} {220403} (\bibinfo {year}
  {2011})}\BibitemShut {NoStop}%
\bibitem [{\citenamefont {{Boseggia}}\ \emph {et~al.}()\citenamefont
  {{Boseggia}}, \citenamefont {{Springell}}, \citenamefont {{Walker}},
  \citenamefont {{Boothroyd}}, \citenamefont {{Prabhakaran}}, \citenamefont
  {{Wermeille}}, \citenamefont {{Bouchenoire}}, \citenamefont {{Collins}},\
  and\ \citenamefont {{McMorrow}}}]{Boseggia2012}%
  \BibitemOpen
  \bibfield  {author} {\bibinfo {author} {\bibfnamefont {S.}~\bibnamefont
  {{Boseggia}}}, \bibinfo {author} {\bibfnamefont {R.}~\bibnamefont
  {{Springell}}}, \bibinfo {author} {\bibfnamefont {H.~C.}\ \bibnamefont
  {{Walker}}}, \bibinfo {author} {\bibfnamefont {A.~T.}\ \bibnamefont
  {{Boothroyd}}}, \bibinfo {author} {\bibfnamefont {D.}~\bibnamefont
  {{Prabhakaran}}}, \bibinfo {author} {\bibfnamefont {D.}~\bibnamefont
  {{Wermeille}}}, \bibinfo {author} {\bibfnamefont {L.}~\bibnamefont
  {{Bouchenoire}}}, \bibinfo {author} {\bibfnamefont {S.~P.}\ \bibnamefont
  {{Collins}}}, \ and\ \bibinfo {author} {\bibfnamefont {D.~F.}\ \bibnamefont
  {{McMorrow}}},\ }\href@noop {} {\bibfield  {journal} {\bibinfo  {journal}
  {ArXiv e-prints}\ }}\Eprint {http://arxiv.org/abs/1201.1452}
  {arXiv:1201.1452} \BibitemShut {NoStop}%
\bibitem [{\citenamefont {Nandi}\ \emph {et~al.}(2010)\citenamefont {Nandi},
  \citenamefont {Kim}, \citenamefont {Kreyssig}, \citenamefont {Fernandes},
  \citenamefont {Pratt}, \citenamefont {Thaler}, \citenamefont {Ni},
  \citenamefont {Bud'ko}, \citenamefont {Canfield}, \citenamefont {Schmalian},
  \citenamefont {McQueeney},\ and\ \citenamefont {Goldman}}]{Nandi2010}%
  \BibitemOpen
  \bibfield  {author} {\bibinfo {author} {\bibfnamefont {S.}~\bibnamefont
  {Nandi}}, \bibinfo {author} {\bibfnamefont {M.~G.}\ \bibnamefont {Kim}},
  \bibinfo {author} {\bibfnamefont {A.}~\bibnamefont {Kreyssig}}, \bibinfo
  {author} {\bibfnamefont {R.~M.}\ \bibnamefont {Fernandes}}, \bibinfo {author}
  {\bibfnamefont {D.~K.}\ \bibnamefont {Pratt}}, \bibinfo {author}
  {\bibfnamefont {A.}~\bibnamefont {Thaler}}, \bibinfo {author} {\bibfnamefont
  {N.}~\bibnamefont {Ni}}, \bibinfo {author} {\bibfnamefont {S.~L.}\
  \bibnamefont {Bud'ko}}, \bibinfo {author} {\bibfnamefont {P.~C.}\
  \bibnamefont {Canfield}}, \bibinfo {author} {\bibfnamefont {J.}~\bibnamefont
  {Schmalian}}, \bibinfo {author} {\bibfnamefont {R.~J.}\ \bibnamefont
  {McQueeney}}, \ and\ \bibinfo {author} {\bibfnamefont {A.~I.}\ \bibnamefont
  {Goldman}},\ }\href {\doibase 10.1103/PhysRevLett.104.057006} {\bibfield
  {journal} {\bibinfo  {journal} {Phys. Rev. Lett.}\ }\textbf {\bibinfo
  {volume} {104}},\ \bibinfo {pages} {057006} (\bibinfo {year}
  {2010})}\BibitemShut {NoStop}%
\bibitem [{\citenamefont {Pratt}\ \emph {et~al.}(2009)\citenamefont {Pratt},
  \citenamefont {Tian}, \citenamefont {Kreyssig}, \citenamefont {Zarestky},
  \citenamefont {Nandi}, \citenamefont {Ni}, \citenamefont {Bud'ko},
  \citenamefont {Canfield}, \citenamefont {Goldman},\ and\ \citenamefont
  {McQueeney}}]{Pratt2009}%
  \BibitemOpen
  \bibfield  {author} {\bibinfo {author} {\bibfnamefont {D.~K.}\ \bibnamefont
  {Pratt}}, \bibinfo {author} {\bibfnamefont {W.}~\bibnamefont {Tian}},
  \bibinfo {author} {\bibfnamefont {A.}~\bibnamefont {Kreyssig}}, \bibinfo
  {author} {\bibfnamefont {J.~L.}\ \bibnamefont {Zarestky}}, \bibinfo {author}
  {\bibfnamefont {S.}~\bibnamefont {Nandi}}, \bibinfo {author} {\bibfnamefont
  {N.}~\bibnamefont {Ni}}, \bibinfo {author} {\bibfnamefont {S.~L.}\
  \bibnamefont {Bud'ko}}, \bibinfo {author} {\bibfnamefont {P.~C.}\
  \bibnamefont {Canfield}}, \bibinfo {author} {\bibfnamefont {A.~I.}\
  \bibnamefont {Goldman}}, \ and\ \bibinfo {author} {\bibfnamefont {R.~J.}\
  \bibnamefont {McQueeney}},\ }\href {\doibase 10.1103/PhysRevLett.103.087001}
  {\bibfield  {journal} {\bibinfo  {journal} {Phys. Rev. Lett.}\ }\textbf
  {\bibinfo {volume} {103}},\ \bibinfo {pages} {087001} (\bibinfo {year}
  {2009})}\BibitemShut {NoStop}%
\bibitem [{Note1()}]{Note1}%
  \BibitemOpen
  \bibinfo {note} {M.G. Kim \protect \emph {et al.} unpublished}\BibitemShut
  {NoStop}%
\bibitem [{\citenamefont {Julien}\ \emph {et~al.}(2009)\citenamefont {Julien},
  \citenamefont {Mayaffre}, \citenamefont {Horvatic}, \citenamefont {Berthier},
  \citenamefont {Zhang}, \citenamefont {Wu}, \citenamefont {Chen},
  \citenamefont {Wang},\ and\ \citenamefont {Luo}}]{Julien2009}%
  \BibitemOpen
  \bibfield  {author} {\bibinfo {author} {\bibfnamefont {M.-H.}\ \bibnamefont
  {Julien}}, \bibinfo {author} {\bibfnamefont {H.}~\bibnamefont {Mayaffre}},
  \bibinfo {author} {\bibfnamefont {M.}~\bibnamefont {Horvatic}}, \bibinfo
  {author} {\bibfnamefont {C.}~\bibnamefont {Berthier}}, \bibinfo {author}
  {\bibfnamefont {X.~D.}\ \bibnamefont {Zhang}}, \bibinfo {author}
  {\bibfnamefont {W.}~\bibnamefont {Wu}}, \bibinfo {author} {\bibfnamefont
  {G.~F.}\ \bibnamefont {Chen}}, \bibinfo {author} {\bibfnamefont {N.~L.}\
  \bibnamefont {Wang}}, \ and\ \bibinfo {author} {\bibfnamefont {J.~L.}\
  \bibnamefont {Luo}},\ }\href {http://stacks.iop.org/0295-5075/87/i=3/a=37001}
  {\bibfield  {journal} {\bibinfo  {journal} {Europhysics Lett.}\ }\textbf
  {\bibinfo {volume} {87}},\ \bibinfo {pages} {37001} (\bibinfo {year}
  {2009})}\BibitemShut {NoStop}%
\bibitem [{\citenamefont {Park}\ \emph {et~al.}(2009)\citenamefont {Park},
  \citenamefont {Inosov}, \citenamefont {Niedermayer}, \citenamefont {Sun},
  \citenamefont {Haug}, \citenamefont {Christensen}, \citenamefont {Dinnebier},
  \citenamefont {Boris}, \citenamefont {Drew}, \citenamefont {Schulz},
  \citenamefont {Shapoval}, \citenamefont {Wolff}, \citenamefont {Neu},
  \citenamefont {Yang}, \citenamefont {Lin}, \citenamefont {Keimer},\ and\
  \citenamefont {Hinkov}}]{Park2009}%
  \BibitemOpen
  \bibfield  {author} {\bibinfo {author} {\bibfnamefont {J.~T.}\ \bibnamefont
  {Park}}, \bibinfo {author} {\bibfnamefont {D.~S.}\ \bibnamefont {Inosov}},
  \bibinfo {author} {\bibfnamefont {C.}~\bibnamefont {Niedermayer}}, \bibinfo
  {author} {\bibfnamefont {G.~L.}\ \bibnamefont {Sun}}, \bibinfo {author}
  {\bibfnamefont {D.}~\bibnamefont {Haug}}, \bibinfo {author} {\bibfnamefont
  {N.~B.}\ \bibnamefont {Christensen}}, \bibinfo {author} {\bibfnamefont
  {R.}~\bibnamefont {Dinnebier}}, \bibinfo {author} {\bibfnamefont {A.~V.}\
  \bibnamefont {Boris}}, \bibinfo {author} {\bibfnamefont {A.~J.}\ \bibnamefont
  {Drew}}, \bibinfo {author} {\bibfnamefont {L.}~\bibnamefont {Schulz}},
  \bibinfo {author} {\bibfnamefont {T.}~\bibnamefont {Shapoval}}, \bibinfo
  {author} {\bibfnamefont {U.}~\bibnamefont {Wolff}}, \bibinfo {author}
  {\bibfnamefont {V.}~\bibnamefont {Neu}}, \bibinfo {author} {\bibfnamefont
  {X.}~\bibnamefont {Yang}}, \bibinfo {author} {\bibfnamefont {C.~T.}\
  \bibnamefont {Lin}}, \bibinfo {author} {\bibfnamefont {B.}~\bibnamefont
  {Keimer}}, \ and\ \bibinfo {author} {\bibfnamefont {V.}~\bibnamefont
  {Hinkov}},\ }\href {\doibase 10.1103/PhysRevLett.102.117006} {\bibfield
  {journal} {\bibinfo  {journal} {Phys. Rev. Lett.}\ }\textbf {\bibinfo
  {volume} {102}},\ \bibinfo {pages} {117006} (\bibinfo {year}
  {2009})}\BibitemShut {NoStop}%
\bibitem [{\citenamefont {Wang}\ \emph {et~al.}(2011)\citenamefont {Wang},
  \citenamefont {Stadnik}, \citenamefont {\ifmmode~\dot{Z}\else
  \.{Z}\fi{}ukrowski}, \citenamefont {Thaler}, \citenamefont {Bud'ko},\ and\
  \citenamefont {Canfield}}]{Wang2011}%
  \BibitemOpen
  \bibfield  {author} {\bibinfo {author} {\bibfnamefont {P.}~\bibnamefont
  {Wang}}, \bibinfo {author} {\bibfnamefont {Z.~M.}\ \bibnamefont {Stadnik}},
  \bibinfo {author} {\bibfnamefont {J.}~\bibnamefont {\ifmmode~\dot{Z}\else
  \.{Z}\fi{}ukrowski}}, \bibinfo {author} {\bibfnamefont {A.}~\bibnamefont
  {Thaler}}, \bibinfo {author} {\bibfnamefont {S.~L.}\ \bibnamefont {Bud'ko}},
  \ and\ \bibinfo {author} {\bibfnamefont {P.~C.}\ \bibnamefont {Canfield}},\
  }\href {\doibase 10.1103/PhysRevB.84.024509} {\bibfield  {journal} {\bibinfo
  {journal} {Phys. Rev. B}\ }\textbf {\bibinfo {volume} {84}},\ \bibinfo
  {pages} {024509} (\bibinfo {year} {2011})}\BibitemShut {NoStop}%
\bibitem [{\citenamefont {Fukazawa}\ \emph {et~al.}(2009)\citenamefont
  {Fukazawa}, \citenamefont {Yamazaki}, \citenamefont {Kondo}, \citenamefont
  {Kohori}, \citenamefont {Takeshita}, \citenamefont {Shirage}, \citenamefont
  {Kihou}, \citenamefont {Miyazawa}, \citenamefont {Kito}, \citenamefont
  {Eisaki},\ and\ \citenamefont {Iyo}}]{Fukazawa2009}%
  \BibitemOpen
  \bibfield  {author} {\bibinfo {author} {\bibfnamefont {H.}~\bibnamefont
  {Fukazawa}}, \bibinfo {author} {\bibfnamefont {T.}~\bibnamefont {Yamazaki}},
  \bibinfo {author} {\bibfnamefont {K.}~\bibnamefont {Kondo}}, \bibinfo
  {author} {\bibfnamefont {Y.}~\bibnamefont {Kohori}}, \bibinfo {author}
  {\bibfnamefont {N.}~\bibnamefont {Takeshita}}, \bibinfo {author}
  {\bibfnamefont {P.~M.}\ \bibnamefont {Shirage}}, \bibinfo {author}
  {\bibfnamefont {K.}~\bibnamefont {Kihou}}, \bibinfo {author} {\bibfnamefont
  {K.}~\bibnamefont {Miyazawa}}, \bibinfo {author} {\bibfnamefont
  {H.}~\bibnamefont {Kito}}, \bibinfo {author} {\bibfnamefont {H.}~\bibnamefont
  {Eisaki}}, \ and\ \bibinfo {author} {\bibfnamefont {A.}~\bibnamefont {Iyo}},\
  }\href {\doibase 10.1143/JPSJ.78.033704} {\bibfield  {journal} {\bibinfo
  {journal} {Journal of the Physical Society of Japan}\ }\textbf {\bibinfo
  {volume} {78}},\ \bibinfo {pages} {033704} (\bibinfo {year}
  {2009})}\BibitemShut {NoStop}%
\bibitem [{\citenamefont {Goko}\ \emph {et~al.}(2009)\citenamefont {Goko},
  \citenamefont {Aczel}, \citenamefont {Baggio-Saitovitch}, \citenamefont
  {Bud'ko}, \citenamefont {Canfield}, \citenamefont {Carlo}, \citenamefont
  {Chen}, \citenamefont {Dai}, \citenamefont {Hamann}, \citenamefont {Hu},
  \citenamefont {Kageyama}, \citenamefont {Luke}, \citenamefont {Luo},
  \citenamefont {Nachumi}, \citenamefont {Ni}, \citenamefont {Reznik},
  \citenamefont {Sanchez-Candela}, \citenamefont {Savici}, \citenamefont
  {Sikes}, \citenamefont {Wang}, \citenamefont {Wiebe}, \citenamefont
  {Williams}, \citenamefont {Yamamoto}, \citenamefont {Yu},\ and\ \citenamefont
  {Uemura}}]{Goko2009}%
  \BibitemOpen
  \bibfield  {author} {\bibinfo {author} {\bibfnamefont {T.}~\bibnamefont
  {Goko}}, \bibinfo {author} {\bibfnamefont {A.~A.}\ \bibnamefont {Aczel}},
  \bibinfo {author} {\bibfnamefont {E.}~\bibnamefont {Baggio-Saitovitch}},
  \bibinfo {author} {\bibfnamefont {S.~L.}\ \bibnamefont {Bud'ko}}, \bibinfo
  {author} {\bibfnamefont {P.~C.}\ \bibnamefont {Canfield}}, \bibinfo {author}
  {\bibfnamefont {J.~P.}\ \bibnamefont {Carlo}}, \bibinfo {author}
  {\bibfnamefont {G.~F.}\ \bibnamefont {Chen}}, \bibinfo {author}
  {\bibfnamefont {P.}~\bibnamefont {Dai}}, \bibinfo {author} {\bibfnamefont
  {A.~C.}\ \bibnamefont {Hamann}}, \bibinfo {author} {\bibfnamefont {W.~Z.}\
  \bibnamefont {Hu}}, \bibinfo {author} {\bibfnamefont {H.}~\bibnamefont
  {Kageyama}}, \bibinfo {author} {\bibfnamefont {G.~M.}\ \bibnamefont {Luke}},
  \bibinfo {author} {\bibfnamefont {J.~L.}\ \bibnamefont {Luo}}, \bibinfo
  {author} {\bibfnamefont {B.}~\bibnamefont {Nachumi}}, \bibinfo {author}
  {\bibfnamefont {N.}~\bibnamefont {Ni}}, \bibinfo {author} {\bibfnamefont
  {D.}~\bibnamefont {Reznik}}, \bibinfo {author} {\bibfnamefont {D.~R.}\
  \bibnamefont {Sanchez-Candela}}, \bibinfo {author} {\bibfnamefont {A.~T.}\
  \bibnamefont {Savici}}, \bibinfo {author} {\bibfnamefont {K.~J.}\
  \bibnamefont {Sikes}}, \bibinfo {author} {\bibfnamefont {N.~L.}\ \bibnamefont
  {Wang}}, \bibinfo {author} {\bibfnamefont {C.~R.}\ \bibnamefont {Wiebe}},
  \bibinfo {author} {\bibfnamefont {T.~J.}\ \bibnamefont {Williams}}, \bibinfo
  {author} {\bibfnamefont {T.}~\bibnamefont {Yamamoto}}, \bibinfo {author}
  {\bibfnamefont {W.}~\bibnamefont {Yu}}, \ and\ \bibinfo {author}
  {\bibfnamefont {Y.~J.}\ \bibnamefont {Uemura}},\ }\href {\doibase
  10.1103/PhysRevB.80.024508} {\bibfield  {journal} {\bibinfo  {journal} {Phys.
  Rev. B}\ }\textbf {\bibinfo {volume} {80}},\ \bibinfo {pages} {024508}
  (\bibinfo {year} {2009})}\BibitemShut {NoStop}%
\bibitem [{\citenamefont {Qi}\ \emph {et~al.}(2011)\citenamefont {Qi},
  \citenamefont {Gao}, \citenamefont {Wang}, \citenamefont {Zhang},
  \citenamefont {Wang}, \citenamefont {Yao}, \citenamefont {Wang},
  \citenamefont {Wang},\ and\ \citenamefont {Ma}}]{Qi2011}%
  \BibitemOpen
  \bibfield  {author} {\bibinfo {author} {\bibfnamefont {Y.}~\bibnamefont
  {Qi}}, \bibinfo {author} {\bibfnamefont {Z.}~\bibnamefont {Gao}}, \bibinfo
  {author} {\bibfnamefont {L.}~\bibnamefont {Wang}}, \bibinfo {author}
  {\bibfnamefont {X.}~\bibnamefont {Zhang}}, \bibinfo {author} {\bibfnamefont
  {D.}~\bibnamefont {Wang}}, \bibinfo {author} {\bibfnamefont {C.}~\bibnamefont
  {Yao}}, \bibinfo {author} {\bibfnamefont {C.}~\bibnamefont {Wang}}, \bibinfo
  {author} {\bibfnamefont {C.}~\bibnamefont {Wang}}, \ and\ \bibinfo {author}
  {\bibfnamefont {Y.}~\bibnamefont {Ma}},\ }\href
  {http://stacks.iop.org/0295-5075/96/i=4/a=47005} {\bibfield  {journal}
  {\bibinfo  {journal} {Europhysics Lett.}\ }\textbf {\bibinfo {volume} {96}},\
  \bibinfo {pages} {47005} (\bibinfo {year} {2011})}\BibitemShut {NoStop}%
\end{thebibliography}
\end{document}